\documentclass[runningheads]{llncs}
\usepackage{graphicx}

\usepackage{tikz}
\usepackage{comment}
\usepackage{amsmath,amssymb} 
\usepackage{color}

\usepackage[accsupp]{axessibility}  

\usepackage{times}

\usepackage{soul}
\usepackage{url}
\usepackage[hidelinks]{hyperref}
\usepackage[utf8]{inputenc}
\usepackage[small]{caption}
\usepackage{subcaption}
\usepackage{graphicx}
\usepackage{amsmath}
\usepackage[numbers]{natbib}
\usepackage{amssymb}
\usepackage{multirow}
\usepackage{booktabs}

\begin{document}
\pagestyle{headings}
\mainmatter
\def\ECCVSubNumber{7754}  

\title{EEG to fMRI Synthesis Benefits from Attentional Graphs of Electrode Relationships} 

\titlerunning{EEG to fMRI Synthesis}
%
\author{David Calhas \and
Rui Henriques}
\authorrunning{D. Calhas et al.}
%
\institute{INESC-ID, Instituto Superior Técnico, Portugal\\
\email{david.calhas@tecnico.ulisboa.pt}}
\maketitle

\begin{abstract}
Topographical structures represent connections between entities and provide a comprehensive design of complex systems. Currently these structures are used to discover correlates of neuronal and haemodynamical activity. In this work, we incorporate them with neural processing techniques to perform regression, using electrophysiological activity to retrieve haemodynamics. To this end, we use Fourier features, attention mechanisms, shared space between modalities and incorporation of style in the latent representation. By combining these techniques, we propose several models that significantly outperform current state-of-the-art of this task in resting state and task-based recording settings. We report which EEG electrodes are the most relevant for the regression task and which relations impacted it the most. In addition, we observe that haemodynamic activity at the scalp, in contrast with sub-cortical regions, is relevant to the learned shared space. Overall, these results suggest that EEG electrode relationships are pivotal to retain information necessary for haemodynamical activity retrieval.
\keywords{Neuroimaging Synthesis, Regression, Machine Learning, Electroencephalography, Functional Magnetic Resonance Imaging}
\end{abstract}


\section{Introduction}\label{section:introduction}

Human brain activity 
upholds cognitive and memory functions. \textit{Neuronal} activity, a set of action potentials localized at the synapses of a neuron \cite{sherrington1952integrative}, can be retrieved through electroencephalography (EEG), while \textit{haemodynamics}, linked to the blood supply \cite{buckner1998event}, measured via functional magnetic resonance imaging (fMRI). Electrophysiological and haemodynamical activity have been widely studied, with several key discoveries on their relationships \cite{shibasaki2008human,yu2016building, he2018spatialtemporaldo, rojas2018study, brechet2019capturing, daly2019electroencephalography,cury2020sparse, abreu2021eeg}.  Though, there is still a research gap in predicting one modality from the other, a problem commonly formulated as a \textit{regression} task also known as  cross-mode mapping or synthesis. The work conducted by \citet{liu2019convolutional} places the state-of-the-art in the synthesis of fMRI from EEG. As simultaneous EEG and fMRI recordings become publicly available \cite{dataset_01, dataset_02, dataset_03}, the neuroscience and machine learning communities can unprecedentedly address these tasks. 
A \textit{synthesized fMRI modality}, sourced only from EEG, promote ambulatory diagnoses, longitudinal monitoring of individuals, and the understanding of synergistic electrophysiological and haemodynamical activity. Given the EEG reduced recording costs 
\cite{de2019introduction}, 
the target synthesis task can further ensure resource availability to communities in need \cite{ogbole2018survey}, providing a proxy to fMRI screens 
with significant impact on the quality of life \cite{van2019value}.

Here, we push the ability of neural processing techniques to perform EEG to fMRI synthesis, and assess what links these two modalities with explainability methods. The gathered results show statistically significant improvements on the target synthesis task against state-of-the-art. The observed breakthroughs are driven by four major and novel contributions: 
\begin{itemize}
    \item conditioning of the fMRI synthesis task with an attention graph that explicitly captures relationships between electrodes;
    \item convolution layering and Fourier feature extraction from EEG-based spectrograms and blood oxygen level dependent (BOLD) signals;
    \item shared latent space between modalities under dedicated losses to aid the training;
    \item incorporation of latent styling principles \cite{gu2021stylenerf} of the attention scores via a \textit{style} posterior on the latent space features.
\end{itemize}


The main findings of this work are:

\begin{itemize}
    \item occipital and parietal relationships with frontal EEG electrodes are the most relevant for haemodynamical retrieval in resting state and temporal electrodes only had relevant relations in task based fMRI synthesis (see Section \ref{section:discussion}). These long topographical links promote Markovian/locality properties to EEG representations (see Section \ref{methods:topo_attention}) and are in accordance with the phenomena of frequencies of the same source being observed in distant electrodes \cite{da2013eeg};
    \item \textit{style} aid in latent representations, a widely used technique in computer vision research \cite{gu2021stylenerf}, is a natural fit when conditioned on these relationships (posterior information)
    , inherently providing robustness for the fMRI synthesis (see Section \ref{results:role_attention_posterior});
    \item results support the claim by \citet{laufs2003eeg, dataset_01} and \citet{dataset_02} on how neuronal activity frequency correlates with fMRI, and further highlight the relevance of spectral features to retrieve haemodynamics             
    (see Section \ref{results:synthesis}).
\end{itemize}

Altogether, these contributions and findings provide a solid ground an EEG to fMRI synthesis framework. To the best of our knowledge, this is the first study to synthesize fMRI using EEG, with real simultaneous EEG and fMRI datasets, whereas the state-of-the-art is applied on synthetic data.

\section{Problem formulation}\label{section:problem_description}
Let $\vec{x} \in \mathbb{R}^{C\times F \times T}$ be an encoding of an EEG recording, where $\vec{x}_i \in \mathbb{R}^{F \times T}$ defines the spectral features of the $i^{th}$ electrode, and $F$ and $T$ correspond to the frequency and temporal dimensions, respectively. Let $\vec{y} \in \mathbb{R}^{V_x \times V_y \times V_z}$ be an fMRI volume representation at a given time, where $V_x$, $V_y$ and $V_z$ correspond to the dimensions across the three dimensional referential axes. Given an arbitrary transformation $f: \mathbb{R}^{C\times F \times T} \to \mathbb{R}^{V_x \times V_y \times V_z}$, the learning objective becomes learning the multi-output regression model, such that $\mbox{argmin}_{\theta} |f(\vec{x}\mid \theta)-\vec{y}|$.

\textcolor{white}{.}

\section{Methods}\label{section:methods}

\begin{figure}[t]
    \centering
    \includegraphics[width=0.7\textwidth]{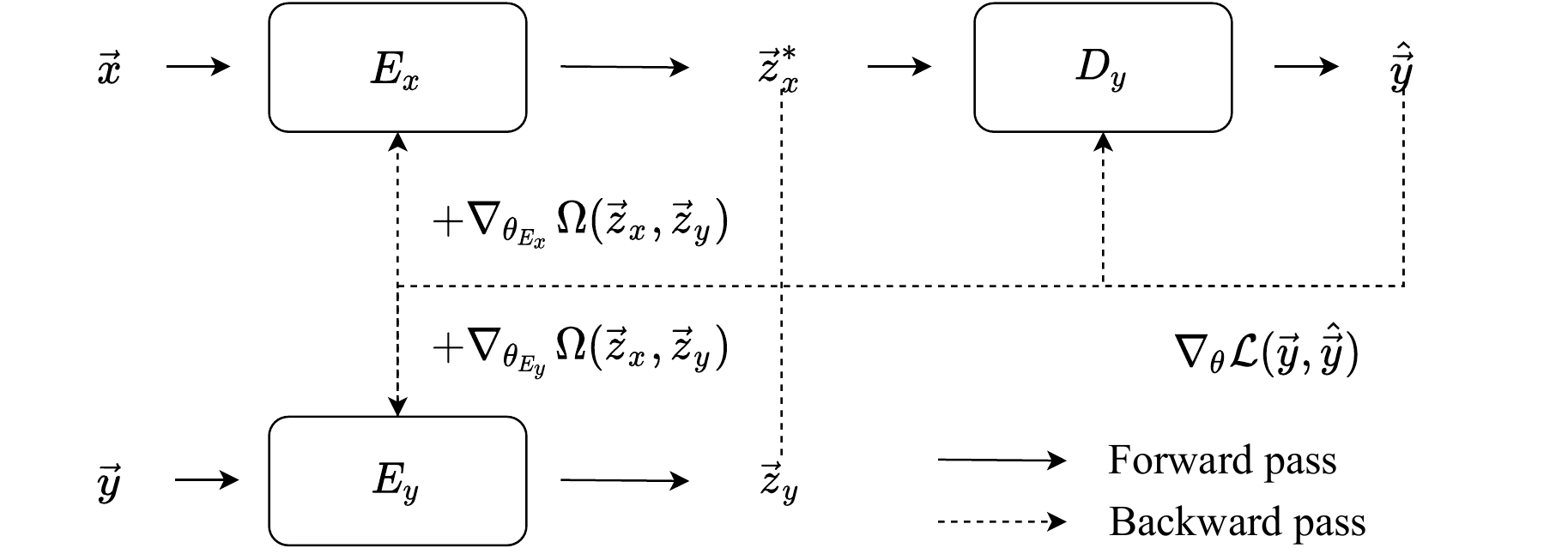}
    \caption{Architectural basis of the EEG to fMRI synthesis with the disclosure of the major forward and backward passes. The fMRI representation, $\vec{y}$, is processed by the encoder, $E_y$, such that $\vec{z}_y = E_y(\vec{y}, \theta_{E_y})$. The EEG representation, $\vec{x}$, in its turn is processed by the encoder, $E_x$, such that $\vec{z}_{x} = E_x(\vec{x}, \theta_{E_x})$. The latent EEG representation, $\vec{z}_{x}$, is processed by the decoder, $D_y$, that performs the mapping to the fMRI estimated instance, $\hat{\vec{y}} = D_y(\vec{x}, \theta_{D_y})$. The gradients for $E_x$, $E_y$ and $D_y$ are $\nabla_{\theta_{E_x}} \mathcal{L}(\vec{y}, \hat{\vec{y}}) + \Omega(\vec{z}_x, \vec{z}_y)$, $\nabla_{\theta_{E_y}} \mathcal{L}(\vec{y}, \hat{\vec{y}}) + \Omega(\vec{z}_x, \vec{z}_y)$ and $\nabla_{\theta_{D_y}} \mathcal{L}(\vec{y}, \hat{\vec{y}})$, respectively.}
    \label{fig:neural_flow}
\end{figure}

Mathematical operations, such as addition and subtraction, are performed over the EEG and fMRI feature representations, $\vec{x}$ and $\vec{y}$, to map the original spaces onto encoded spaces that are identical in structure, $\vec{z}_x$ and $\vec{z}_y$, respectively. This is performed in accordance with the methodology described in Appendix \ref{setting:auto_nas}. 
To this end, architecture modules of Resnet-18 blocks is optimized using Calhas et al. \cite{calhas2022automatic} framework, which hyperparameterizes kernel and stride sizes, potentially differing from layer to layer. The heterogeneity of this convolutional layering structure is beneficial for the performance of the model according to \citet{anonymous2022learning}. 
Following, $\vec{z}_x$ is processed by a densely connected layer with a linear activation to perform the decoding from the encoded representation onto the fMRI volume. 
Figure \ref{fig:neural_flow} illustrates the described neural processing pipeline. 
For the sake of simplicity, please consider $\theta = \theta_{E_y} \bigcup \theta_{E_x} \bigcup \theta_{D_y}$.

\subsection{Latent EEG Fourier features}\label{methods:fourier}

Fourier feature extraction \cite{rahimi2007random} is applied, in addition to spectral analysis, to address the representational dissimilarity gap between EEG and fMRI features spaces. 
The underlying contributions in computer vision \cite{rahimi2007random, tancik2020fourier, gu2021stylenerf} have shown the ability of Fourier features to capture functions with high resolutions. The high degrees of freedom of such an operation \cite{li2019towards} support the synthesis of an fMRI representation with rich spatial resolution. 

Fourier features are applied, as a densely connected layer, to the latent representation $\vec{z}_x \in \mathbb{R}^{L}$ according to
\begin{equation}
    \mbox{cos}(\omega \vec{z}_x + b),
\end{equation}
where $\omega \sim \mathcal{N}(0,1)$ and $b \sim \mathcal{U}(0, 2\pi)$. This layer is initialized with $L$ independently drawn samples from $\mathcal{N}(0,1)$ and $\mathcal{U}(0, 2\pi)$, such that $[\omega_1, \dots ,\omega_L]$ and $[b_1, \dots, b_L]$ form the projections for
\small{
\begin{equation}\label{equation:}
    \vec{z}_{x}^* = \sqrt{\frac{2}{L}} \begin{bmatrix}
    \mbox{cos}(\omega_1 \vec{z}_x + b_1) & \dots & \mbox{cos}(\omega_L \vec{z}_x + b_L)
    \end{bmatrix},
\end{equation}
}
\normalsize
defining the random Fourier features.

\subsection{Topographical attention}\label{methods:topo_attention}

The EEG electrodes are placed on the human scalp in accordance to a certain system, e.g. 10-20 system \cite{jasper1958ten}. It is known that frequencies from the same neuronal source can be present in distant electrodes \cite{da2013eeg} producing an EEG representation without locality, which makes them distant from natural images, a.k.a. Markovian images. \citet{banville2022118994} goes further and claims such a mechanism is robust against ill defined EEG electrodes. This type of schema/relationship between electrodes is not able to be encoded in an Euclidean space, with topographical structures, such as graphs, usually being the go to approach. Since the layers of the encoder, $E_x$, are convolutional layers, relying on the Markovian property of its inputs \cite{krizhevsky2012imagenet}, one needs to promote locality with a reordering operation. To that end, we propose the use of an attention mechanism at the level of the EEG electrodes. Let $A \in \mathbb{R}^{C \times F \times T}$ be the attention weights, such that $\forall i \in \{1, \dots, C\}: A_i \in \mathbb{R}^{F \times T} $, and $E \in \mathbb{R}^{C\times C}$ the context matrix where each column $i \in \{1, \dots, C\}: e_i^\top = \begin{bmatrix} e_{i1} & \dots & e_{iC} \end{bmatrix} = \begin{bmatrix} \vec{x}_i^\top A_{1} & \dots & \vec{x}_i^\top A_{C} \end{bmatrix}$. The attention scores, 
\begin{equation}\label{equation:attention_scores}
W = \begin{bmatrix} w_{11} & \dots & w_{1C}\\
                    \vdots & \ddots & \vdots\\
                    w_{C1} & \dots & w_{CC} \end{bmatrix}, \hspace{0.2cm}\text{where}\hspace{0.2cm} w_i^\top = \begin{bmatrix} \frac{e^{e_{1}}}{\sum_j e^{e_{j}}} & \dots & \frac{e^{e_{C}}}{\sum_j e^{e_{j}}} \end{bmatrix},
\end{equation}
are used to produce the output $T \in \mathbb{R}^{C \times F \times T}$, such that
\begin{equation}\label{equation:hadammard_attention}
    T_i = \sum^C \vec{x} \odot w_i.
\end{equation}
The proposed mechanism reorders electrode features to allow locality and performs a element wise (Hadamard) product, denoted by $\odot$, 
on the electrode dimension, $C$. 
\vskip 0.3cm

\noindent\textbf{Adding style conditioned on attention scores.} According to latent style principles, as done by \citet{gu2021stylenerf}, the attention scores, $W$, are used to add \textit{style} features to the latent representation, $z^*_x$. This is done by placing a \textit{style} posterior of the form $z_w|W,\vec{x}: z_w = B^T W$, as

\begin{equation}\label{equation:attention_style_posterior}
    \vec{z}_x^* \odot z_w,
\end{equation}
with $B \in \mathbb{R}^{C \times C \times L}$.

\section{Experimental setting}\label{section:setting}

The mean absolute error (MAE) is used as the loss function, $\mathcal{L}$, and the cosine distance between the latent representations of the EEG and fMRI was added as a regularization term, $\Omega$. MAE is known to be robust under limited data settings and the regularization term on latent representations forces the network to approximate the target representation at the earlier layers \cite{tran2018dist}. The regularization constant was set to $1$ and not optimized for the sake of proof of concept. Please recall the gradient computation illustrated in Figure \ref{fig:neural_flow}. 

\subsection{Bayesian optimization}\label{setting:bayesian_optimization}

The hyperparameters are optimized in two phases:

\begin{enumerate}
    \item using the fMRI autoencoder, $D_y(E_y(.; \theta_{E_y}); \theta_{D_y})$, to discover the optimal dimension of the latent space, $L: E_y(.; \theta_{E_y}) \in \mathbb{R}^{K \times K \times K}$;\footnote{Recall that the latent dimension is defined as both $\mathbb{R}^{K \times K \times K} = \mathbb{R}^{L}$, since $L=K \times K \times K$.}
    \item using the complete neural flow, illustrated in Figure \ref{fig:neural_flow}, that maps EEG to fMRI.
\end{enumerate}

In both phases, Bayesian Optimization \cite{snoek2012practical} is applied with a total of $100$ iterations. The hyperparameters subject to optimization, with their respective domains, are:

\begin{itemize}
    \item learning rate $\in [1\mbox{e}-10, 1\mbox{e}-2]$;
    \item weight decay $\in [1\mbox{e}-10, 1\mbox{e}-1]$;
    \item filter size $\in \{2,4\}$;
    \item max pooling layers (after each convolutional layer) $\in \{0,1\}$;
    \item batch normalization layers (after each max pooling/convoutional layer) $\in \{0,1\}$;
    \item skip layers (in Resnet block) $\in \{0,1\}$;
    \item dropout of convolutional weights $\in \{0,1\}$.
\end{itemize}

In phase 1, the latent dimension is $K \in \{4,5,6,7,8\}$, the max pooling, batch normalization and skip connection layers have a search space of $\{0,1\}$, $0$ means they do not participate in the architecture and $1$ means they do. The batch size was fixed at $64$ to decrease the time spent on the hyperparameter optimization. The optimization was performed on the \citet{dataset_01} dataset, where a total of $8$ individuals and $297$ volumes per individual were considered. A split of $75/25$ was applied to define the training and validation sets.

In addition to the Bayesian optimization, a neural architecture search procedure is also set, as well as an automation of neural architecture generation \cite{calhas2022automatic}, explained in Appendix \ref{setting:auto_nas}. 
Altogether, these steps make the methodology as unbiased as possible from prior and domain based assumptions, removing the human bias and strengthening the generalization.

\subsection{Datasets}\label{setting:datasets}

In total, three simultaneous EEG and fMRI datasets were considered to gather experimental results. In this section, a description of each is provided.
\vskip 0.2cm
\noindent\textbf{NODDI} by \citet{dataset_01}. A dataset with 10 individuals under resting state with eyes open recordings. The EEG recording setup has a total of 64 channels placed according to the 10-20 system, sampled at 250Hz. The fMRI recording was performed with a $2.160$ second Time Response (TR) and $30$ milliseconds echo time (TE). Each voxel is $3 \times 3 \times 3$mm and the resolution of a volume is $64 \times 64 \times 30$. For this dataset 8 individuals were considered for training and 2 individuals used to form the testing set.

\vskip 0.2cm
\noindent\textbf{Oddball} by \citet{dataset_02}. A dataset with 10 individual subjected to a visual object detection task setting. The EEG recording setup has a total of 43 channels, sampled at 1000Hz. The fMRI recording was performed with a $2$ second TR. Each voxel is $3 \times 3 \times 3$mm and the resolution of a volume is $64 \times 64 \times 32$. Similarly, for this dataset 8 individuals were considered for training and 2 individuals for testing.

\vskip 0.2cm
\noindent\textbf{CN-EPFL} by \citet{dataset_03}. A dataset with 20 individuals performing an activity during the recording session. The EEG recording setup has a total of 64 channels placed according to the 10-20 system and sampled at 5000Hz. The fMRI recording was performed with a $1.280$ second TR and $31$ milliseconds TE. Each voxel is $2 \times 2 \times 2$mm and the resolution of a volume is $108 \times 108 \times 54$. Due to the high spatial resolution of this dataset it can be quite memory consuming to run a neural network that performs an affine transformation to a total of $108 \times 108 \times 54$ voxels. To alleviate memory consumption, the discrete cosine transform (DCT II) \cite{ahmed1974discrete} was performed and the fMRI volumes were downsampled to $64 \times 64 \times 30$ voxels, by cutting the frequency coefficients and doing the inverse discrete cosine transform (DCT III). For this dataset 16 individuals were considered for training and 4 individuals composed the testing set.

\vskip 0.2cm
\noindent\textbf{Data preprocessing.} No additional preprocessing was undertaken on the fMRI other than to the preprocessing performed by the corresponding studies linked to the datasets. In contrast, the EEG representations were modified by applying a short-time Fourier transform \cite{allen1977short} with a window of 2 seconds length in order to assess frequencies as low as $0.5$Hz (delta waves). Although lower frequencies are not ranked as the most relevant correlations with haemodynamics \cite{laufs2003eeg}, they are nonetheless informative. Then, the pairing of EEG, $\vec{x}$, and an fMRI volume, $\vec{y}$, was done in such a way that $20$ seconds of EEG were considered for a single fMRI volume. Only EEG information, from before the previous $6$ seconds the fMRI volume was taken. This goes in accordance with the claim that neuronal activity only reflects changes in haemodynamics $5.4$ to $6$ seconds after \cite{liao2002593}. As such, $\vec{x}$ is taken in an interval $[t-26,t-6]$, being $t$ the referenced time when the fMRI volume was taken. Consequently, formalizing the EEG representation $\vec{x} \in \mathbb{R}^{C\times F \times 20}$.

\section{Results}\label{section:results}

Using the introduced experimental setting, results are produced under the methodology described in Section \ref{section:methods} and assessed against the state-of-the-art approaches by \citet{liu2019convolutional}.\footnote{Please note that the baseline of \cite{liu2019convolutional} is implemented by the authors as there is no public implementation by the original study. Further, only the EEG to fMRI description was considered and only one volume is synthesized, as opposed to all volumes at once as done in \cite{liu2018darts}. The model was trained to minimize the MAE loss function.} The hyperparameters were optimized in the NODDI dataset and are reported in Table \ref{tab:hyperparameters}. 
These were used for the experiments of all the other datasets.

The baselines subject to comparison with the state-of-the-art are:

\begin{itemize}
    \item (i) Linear projection on the latent space representation, $\vec{z}_x$;
    \item (ii) [with \textit{style} posterior] Topographical attention on the EEG electrode dimension;
    \item (iii) Random Fourier feature \cite{tancik2020fourier} projection on the latent space representation, $\vec{z}_x^*$;
    \item (iv) [with \textit{style} posterior] Combination of (ii) and (iii), as topographical attention is applied in the EEG electrode dimension, as well as the random Fourier feature \cite{tancik2020fourier} projection on the latent space representation, $\vec{z}_x^*$.
\end{itemize}

Additionally, experiments of (ii) and (iv), with no style and with a \textit{style} prior learnable vector, are reported in section \ref{results:role_attention_posterior}.

\subsection{fMRI synthesis}\label{results:synthesis}

\begin{table*}[t]
    \centering
    \tiny{
    \begin{tabular}{ p{1.75cm} | p{1.6cm} p{1.6cm} p{1.6cm} | p{1.6cm} p{1.6cm} p{1.6cm}}
        \hline
        \hline
        \hfil \multirow{2}{*}{Model}  & \multicolumn{3}{c}{RMSE} &  \multicolumn{3}{|c}{SSIM}\\
         & \hfil NODDI & \hfil Oddball  & \hfil CN-EPFL   & \hfil NODDI & \hfil Oddball  & \hfil CN-EPFL    \\
         \hline
        \hfil (i) & \hfil $0.5124 \pm 0.0498$ & \hfil $0.7419 \pm 0.0290$ & \hfil $0.5860 \pm 0.0865$ & \hfil $0.4329 \pm 0.0054$ & \hfil $0.1829 \pm 0.0332$ & \hfil $0.5037 \pm 0.0734$ \\
        \hfil (ii) (with \textit{style} posterior) & \hfil $0.4121 \pm 0.0390$ & \hfil $0.7728 \pm 0.1184$ & \hfil $0.5288 \pm 0.0355$ & \hfil $0.4724 \pm 0.0096$ & \hfil $0.1580 \pm 0.0405$ & \hfil $0.5221 \pm 0.0707$\\
        \hfil (iii) & \hfil $0.4333 \pm 0.0448$ & \hfil $0.7326 \pm 0.0463$ & \hfil $0.5282 \pm 0.0614$ & \hfil $0.4618 \pm 0.0028$ & \hfil $0.1963 \pm 0.0388$ & \hfil $0.5074 \pm 0.0833$\\
        \hfil (iv) (with \textit{style} posterior) & \hfil $0.3972 \pm 0.0186$ & \hfil $0.7014 \pm 0.0855$ & \hfil $0.5166 \pm 0.0560$ & \hfil $0.4613 \pm 0.0198$ & \hfil $0.2004 \pm 0.0172$ & \hfil $0.5222 \pm 0.0877$\\
        \hline
        \hfil \citet{liu2019convolutional} & \hfil $0.4549 \pm 0.0806$ & \hfil $0.8591 \pm 0.0342$ & \hfil $0.5915 \pm 0.1083$ & \hfil $0.4488 \pm 0.0601$ & \hfil $0.1885 \pm 0.0380$ & \hfil $0.5190 \pm 0.1062$\\
        \hline
        \hline
    \end{tabular}}
    \caption{Root mean squared error (RMSE) and structural similarity index measure (SSIM) of the target synthesis task for the proposed and state-of-the-art models across all datasets. 
    (i) refers to the linear projection in the latent space, (ii) refers to topographical attention on the EEG channels dimensions with a linear projection in the latent space, (iii) implements a random Fourier feature projection in the latent space, and (iv) performs topographical attention on the EEG channels dimension with a random Fourier features projection in the latent space.}
    \label{tab:quantitative}
\end{table*}

\begin{figure*}[t]
     \centering
     \begin{subfigure}[t]{0.48\textwidth}
         \centering
         \includegraphics[width=\textwidth]{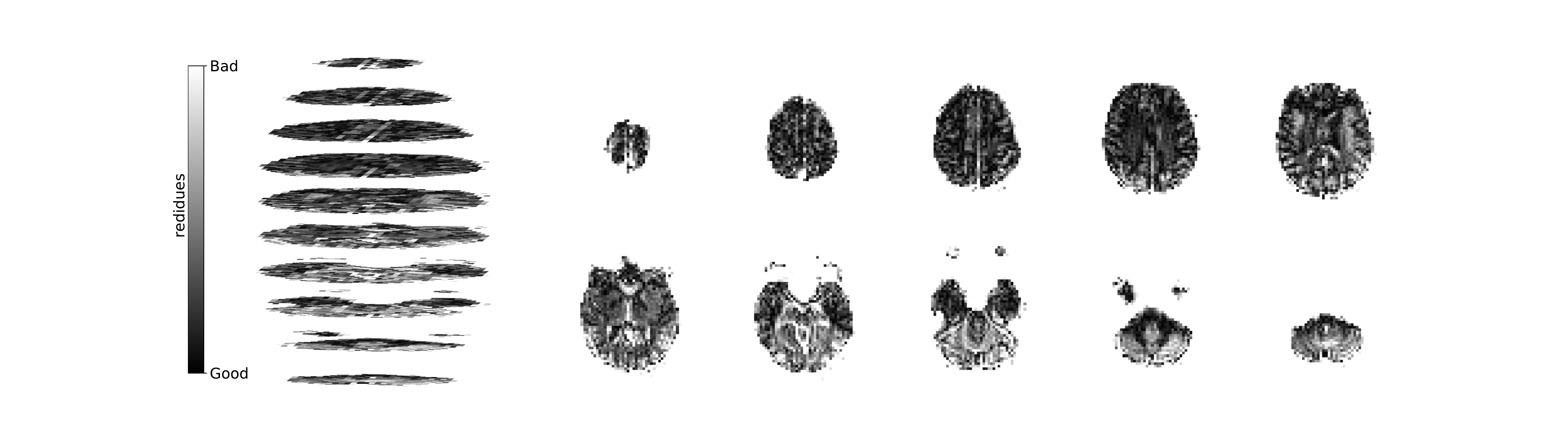}
         \caption{(i) - Linear latent projection. RMSE of $0.5124 \pm 0.0498$ and SSIM of $0.4329 \pm 0.0054$.}
         \label{fig:01_linear_mean_residues}
     \end{subfigure}
     \hfill
     \begin{subfigure}[t]{0.48\textwidth}
         \centering
         \includegraphics[width=\textwidth]{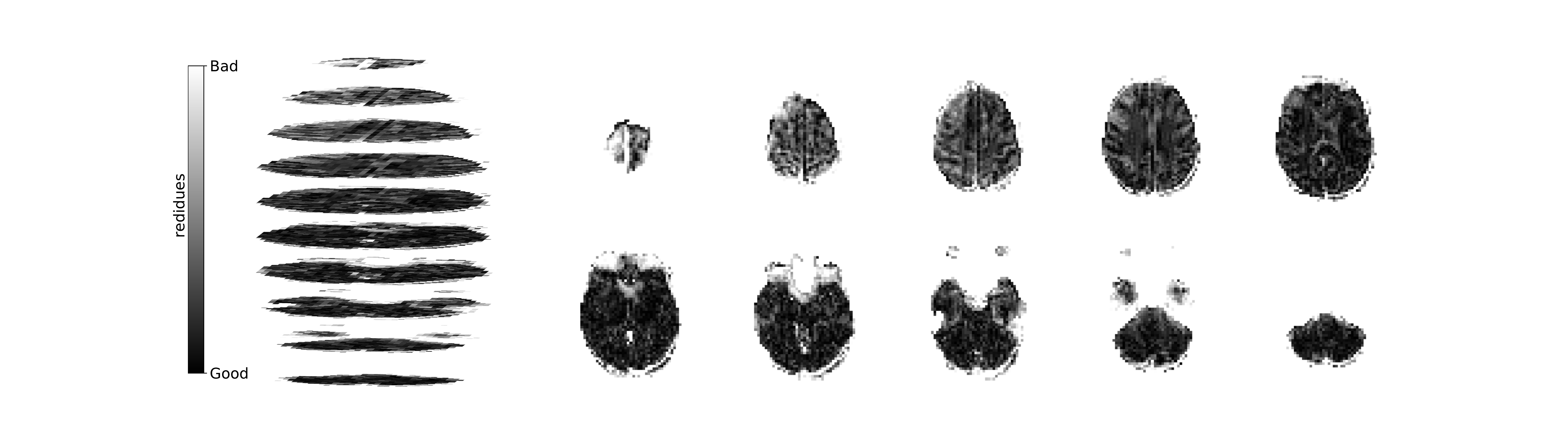}
         \caption{(ii) - Topographical attention on the EEG channels dimension, with linear latent projection (i). Attention scores are placed as a \textit{style} posterior on the latent representation, as described in Equation \ref{equation:attention_style_posterior}. RMSE of $0.4121 \pm 0.0390$ and SSIM of $0.4724 \pm 0.0096$.}
         \label{fig:01_topo_mean_residues}
     \end{subfigure}
     ~
     \begin{subfigure}[t]{0.48\textwidth}
         \centering
         \includegraphics[width=\textwidth]{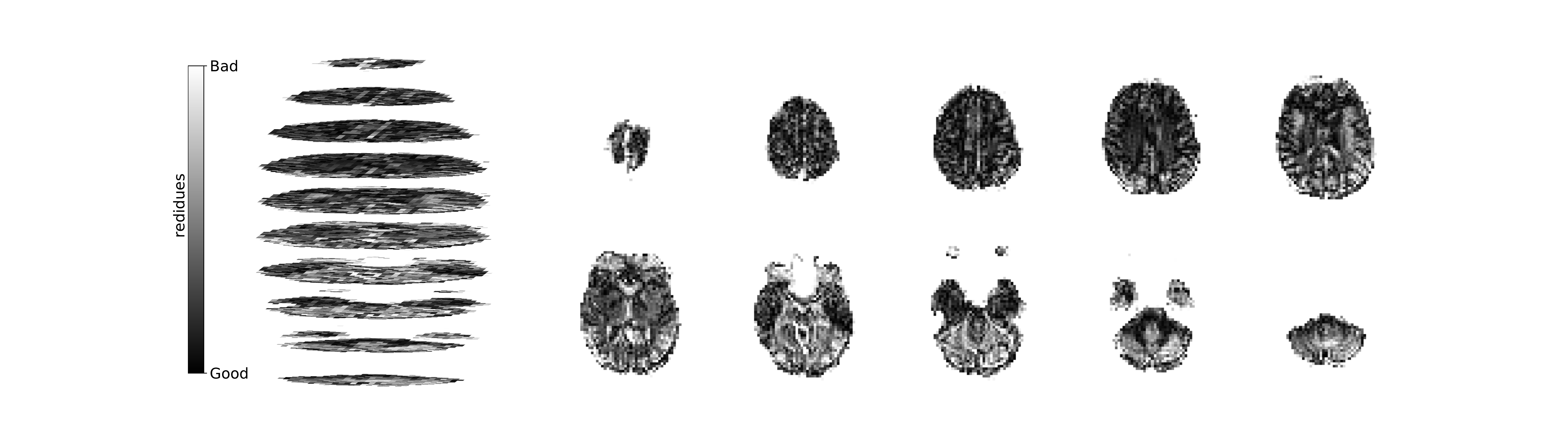}
         \caption{(iii) - Random Fourier feature latent projection. RMSE of $0.4333 \pm 0.0448$ and SSIM of $0.4618 \pm 0.0028$.}
         \label{fig:01_fourier_mean_residues}
     \end{subfigure}
     \hfill
     \begin{subfigure}[t]{0.48\textwidth}
         \centering
         \includegraphics[width=\textwidth]{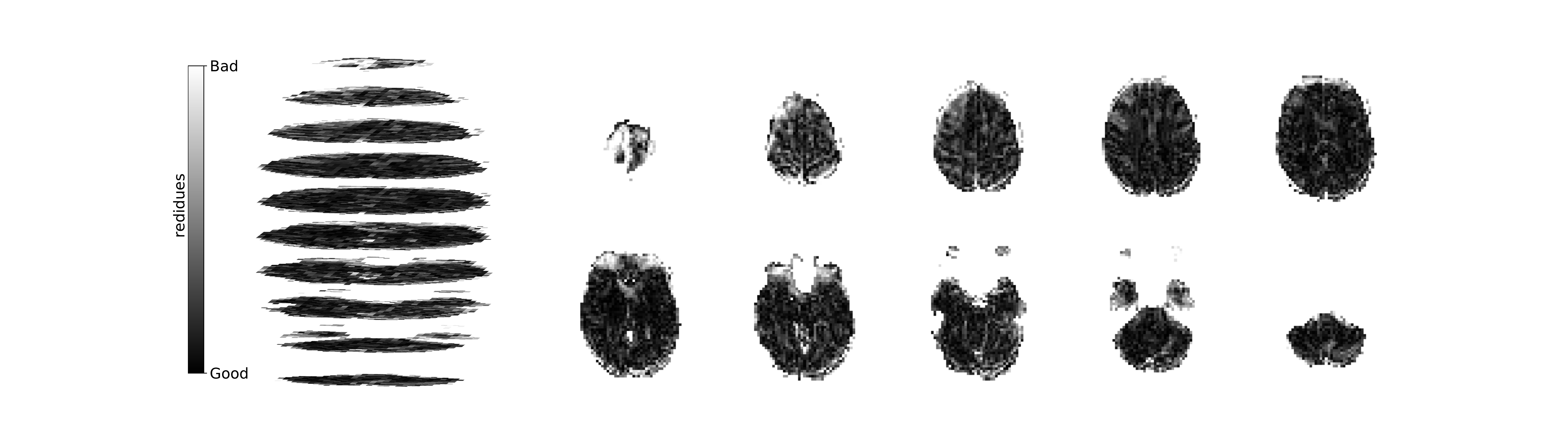}
         \caption{(iii) - Topographical attention on the EEG channels dimension, with random Fourier latent projection (i). Attention scores are placed as a \textit{style} posterior on the latent representation, as described in Equation \ref{equation:attention_style_posterior}. RMSE of $0.3972 \pm 0.0186$ and SSIM of $0.4613 \pm 0.0198$.}
         \label{fig:01_fourier_topo_mean_residues}
     \end{subfigure}
    \caption{Mean absolute residues for each implemented models. 
    Model (ii), implementing topographical attention with a \textit{style} posterior, and model (iv), additionally transforming the latent features using the random Fourier feature projection (described in Section \ref{methods:fourier}), achieve the best performance relative to RMSE and SSIM metrics.}
    \label{fig:mean_residues}
\end{figure*}

Figure \ref{fig:mean_residues} illustrates the distribution of residues (observed vs. estimated differences) on the fMRI volumes for the NODDI dataset. Clearly, by visual inspection, (iv) model has the darker and biggest area of shaded regions, which implies a better coverage across the brain regions and better synthesis quality. Models with topographical attention, (ii) and (iv), corresponding to Figures \ref{fig:01_topo_mean_residues} and \ref{fig:01_fourier_topo_mean_residues}, respectively, significantly improve the synthesis, as shown by the darker and bigger areas against (i) and (iii) depicted in Figures \ref{fig:01_linear_mean_residues} and \ref{fig:01_fourier_mean_residues}, respectively. Particularly, we notice that models (i) and (iii) report difficulty in the retrieval of haemodynamical activity located in occipital and parietal lobes.

\begin{figure*}[t]
     \centering
     \begin{subfigure}[t]{0.48\textwidth}
         \centering
         \includegraphics[width=\textwidth]{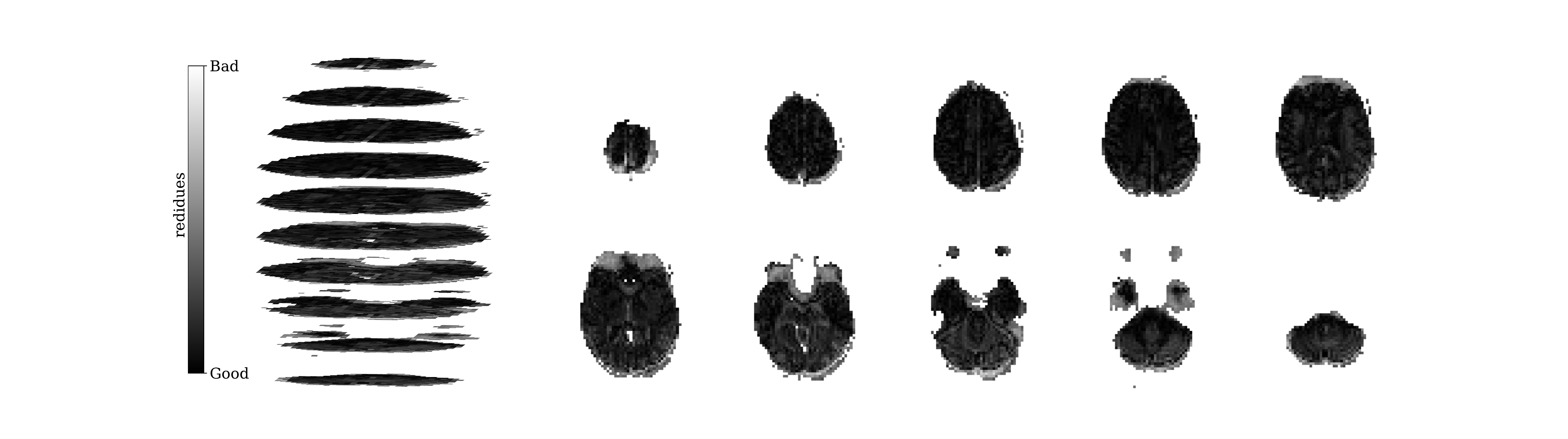}
         \caption{(i) - Linear latent projection.}
         \label{fig:01_linear_norm_mean_residues}
     \end{subfigure}
     \hfill
     \begin{subfigure}[t]{0.48\textwidth}
         \centering
         \includegraphics[width=\textwidth]{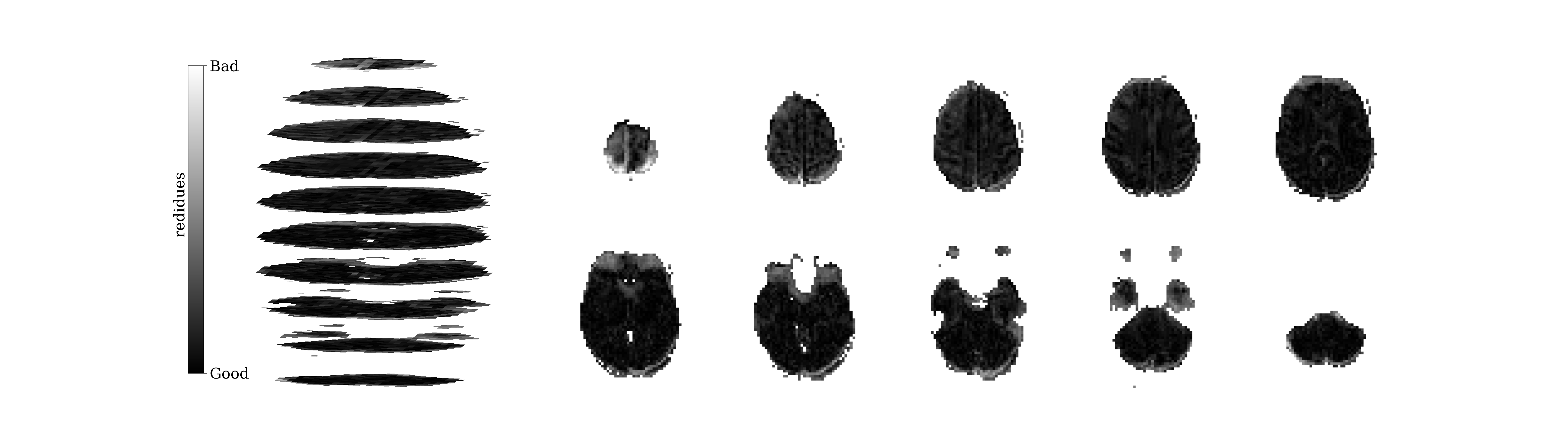}
         \caption{(ii) - Topographical attention on the EEG channels dimension, with linear latent projection (i).}
         \label{fig:01_topo_norm_mean_residues}
     \end{subfigure}
     ~
     \begin{subfigure}[t]{0.48\textwidth}
         \centering
         \includegraphics[width=\textwidth]{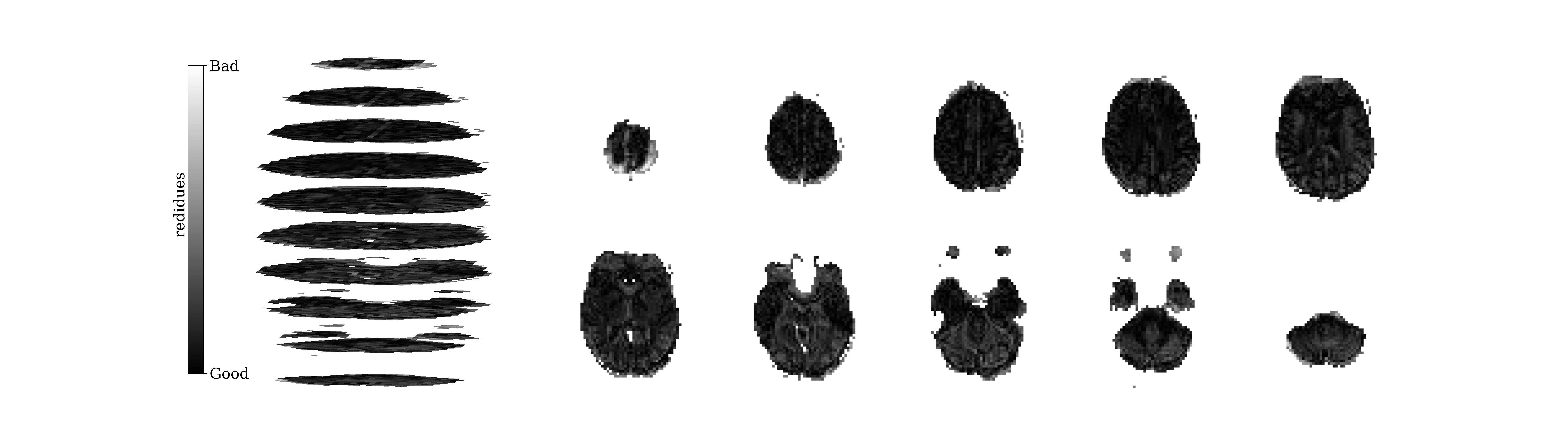}
         \caption{(iii) - Random Fourier feature latent projection.}
         \label{fig:01_fourier_norm_mean_residues}
     \end{subfigure}
     \hfill
     \begin{subfigure}[t]{0.48\textwidth}
         \centering
         \includegraphics[width=\textwidth]{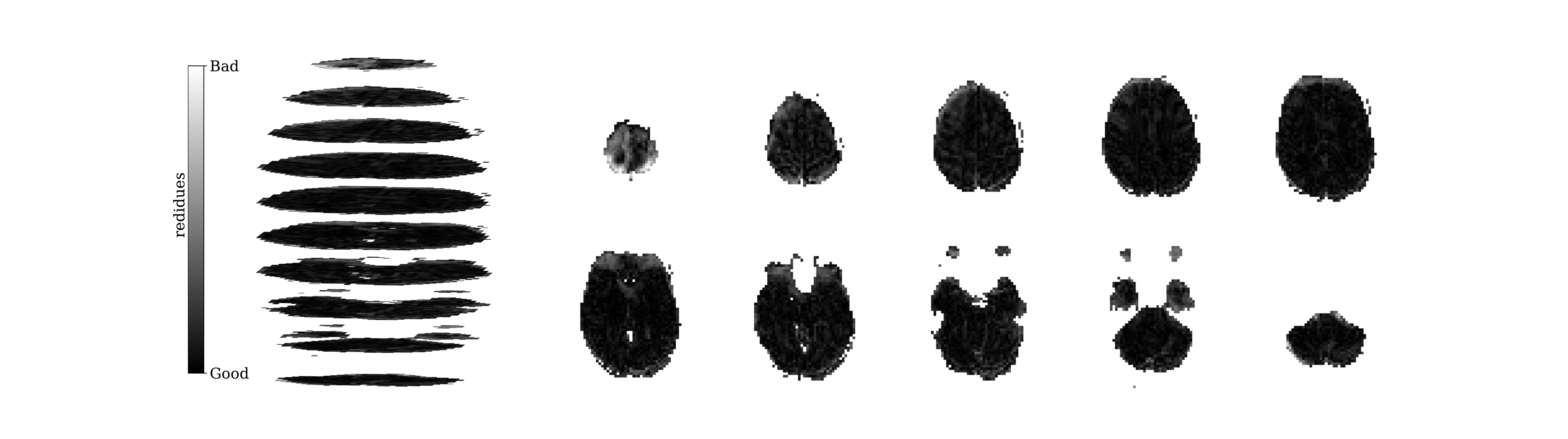}
         \caption{(iii) - Topographical attention on the EEG channels dimension, with random Fourier latent projection (i).} 
         \label{fig:01_fourier_topo_norm_mean_residues}
     \end{subfigure}
    \caption{Normalized mean absolute residues for the proposed models.}
    \label{fig:mean_normalized_residues}
\end{figure*}

To better address which regions our baselines had more difficulty retrieving, the normalized residues were computed and are illustrated in Figure \ref{fig:mean_normalized_residues}. Baselines -- corresponding to models (i) and (ii), shown in Figures \ref{fig:01_linear_norm_mean_residues} and \ref{fig:01_topo_norm_mean_residues} respectively, which correspondingly implement a linear projection in the latent space and topographical attention --, have difficulty retrieving the prefrontal, occipital and parietal lobes, as the shade tends to a lighter grey in that region. Model (iv), shown in Figure \ref{fig:01_fourier_topo_norm_mean_residues}, does not show a noticeable region with a lighter tone of grey, which implies no evident difficulty in retrieving haemodynamical activity across the different brain regions.

Table \ref{tab:quantitative} contains the results obtained from running the target approaches ((i), (ii), (iii) and (iv)) and the state-of-the-art \cite{liu2019convolutional}. For all datasets considered in the experiments, model (iv) obtained the best RMSE values. Further, our baselines consistently outperform the state-of-the-art, according to the RMSE metric. From analyzing our baselines, we conclude that random Fourier features, described in section \ref{methods:fourier}, benefit models (i) and (ii) and the introduction of topographical attention also benefits both models (i) and (iii). The latter, shows the adaptability and robustness of introducing topographical relationships to the synthesis of fMRI. By assessing the experiments from the perspective of the SSIM metric, there is not a concordant superiority across all datasets, as observed with the RMSE. Nonetheless, the state-of-the-art is outperformed by at least one of our baselines on all datasets. Specifically, on the NODDI dataset (resting state), we observe that incorporation of topographical attention in model (ii), under a \textit{style} posterior, achieves the best SSIM value. 

\begin{figure*}[t]
    \centering
    \includegraphics[clip, trim=5cm 9cm 5cm 13.5cm,width=\textwidth]{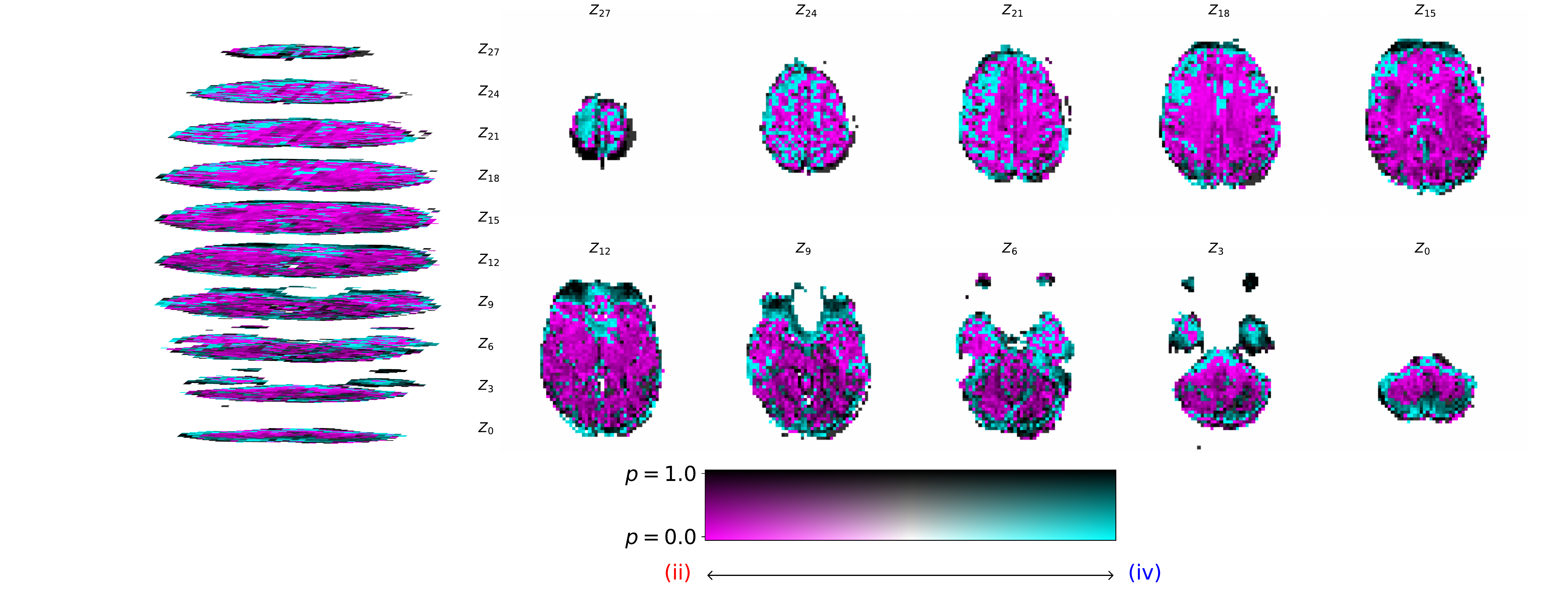}
    \caption{Region-sensitive comparison of models (ii) and (iv), both using \textit{style} posterior, reporting the best model in each voxel according to predictive power (statistical significance under $t$-test). Although Table \ref{tab:quantitative} shows that model (iv) outperforms (ii) regarding RMSE, this analysis shows that model (ii) achieves a significantly better synthesis capacity on the majority of the voxels.}
    \label{fig:01_ii_vs_iv}
\end{figure*}

Figure \ref{fig:01_ii_vs_iv} illustrates the voxel wise comparance, with statistical significance report, between (ii) and (iv). Figure \ref{fig:comparison_plots} reports the same comparison for the rest of the models in NODDI dataset.

\begin{table*}[t]
    \centering
    \tiny{
    \begin{tabular}{ p{1.75cm} | p{1.6cm} p{1.6cm} p{1.6cm} | p{1.6cm} p{1.6cm} p{1.6cm}}
        \hline
        \hline
        \hfil \multirow{2}{*}{}  & \multicolumn{3}{c}{RMSE} &  \multicolumn{3}{|c}{SSIM}\\
         & \hfil NODDI & \hfil Oddball & \hfil CN-EPFL & \hfil NODDI & \hfil Oddball & \hfil CN-EPFL  \\
         \hline
        \hfil (ii) w/o \textit{style} & \hfil $0.5119 \pm 0.0494$ & \hfil $0.9812 \pm 0.0847$ & \hfil $0.5458 \pm 0.0596$ & \hfil $0.4322 \pm 0.0054$ & \hfil $0.1930 \pm 0.0543$ & \hfil $0.5027 \pm 0.0748$\\
        \hfil (iv) w/o \textit{style} & \hfil $0.4321 \pm 0.0418$ & \hfil $0.7221 \pm 0.0411$ & \hfil $0.5298 \pm 0.0636$ & \hfil $0.4621 \pm 0.0027$ & \hfil $0.1991 \pm 0.0382$ & \hfil $0.5063 \pm 0.0830$ \\
        \hline
        \hfil (ii) (with \textit{style} prior) & \hfil $0.5159 \pm 0.0477$  & \hfil $0.9920 \pm 0.8901$  & \hfil $0.9920 \pm 0.8901$ & \hfil $0.4300 \pm 0.0043$ & \hfil $0.1760 \pm 0.0402$ & \hfil $0.4974 \pm 0.1353$ \\
        \hfil (iv) (with \textit{style} prior) & \hfil $0.4833 \pm 0.0483$ & \hfil $0.7394 \pm 0.0377$  & \hfil $0.5568 \pm 0.0737$ & \hfil $0.4388 \pm 0.0069$ & \hfil $0.1873 \pm 0.0347$ & \hfil $0.4960 \pm 0.1084$\\
        \hline
        \hline
    \end{tabular}}
    \caption{RMSE and SSIM scores in the absence and presence of prior styling, all considering the presence of a posterior style vector conditioned on the attention scores. 
    The upper half of this table shows the results of implementing topographical attention, but without using the attention scores to add style to the latent space representation (w/o style). The bottom half, shows the use of a style prior vector, $\in \mathbb{R}^L$, that is not conditioned on any features, and serves to add learnable style features to the latent representation. The latter is widely used in computer vision research, with a recent study applying it to generate images \cite{gu2021stylenerf}.}
    \label{tab:style}
\end{table*}

For the Oddball dataset, the RMSE and SSIM metrics report a worse synthesis ability for all methodologies compared to the other datasets. Our baselines outperform the state-of-the-art, and model (iv) with a \textit{style} posterior is significantly superior to all baselines. Random Fourier projections, (iii), appear to better address the synthesis task than topographical attention alone, (ii). The SSIM is rather poor, with values below $0.2000$ being the mean and only model (iv) surpassing this threshold with $0.2004$ SSIM.

Models (ii) and (iv), both implementing topographical attention with a \textit{style} posterior, show the best performance in terms of SSIM metric in the CN-EPFL dataset. In spite of the RMSE and SSIM not being in total accordance, the topographical attention superiority is consistent for the metrics considered. This supports our hypothesis that the use of topographical structures plays an important role when studying these two modalities and is hence preferable.

\subsection{Role of topographical attention}\label{results:role_attention_posterior}

In Table \ref{tab:quantitative}, we reported results of models (ii) and (iv), both implementing a \textit{style} posterior vector that is conditioned on the learned attention graph. This graph is a representation of the relationships between the EEG electrodes, learned during the optimization process, that inherently help the retrieval of haemodynamical activity. To validate this hypothesis, Table \ref{tab:style} shows the RMSE and SSIM metrics obtained from experiments ran on the following models:

\begin{itemize}
    \item (ii) and (iv) with no \textit{style} induction, but still performing attention in the EEG electrode dimension;
    \item (ii) and (iv) with \textit{style} prior, reported on the bottom half.
\end{itemize}

From the previous section, we know that the topographical attention, inducing a \textit{style} posterior on the latent representation (see Section \ref{equation:attention_style_posterior}), consistently benefits the regression task across all the datasets considered in our experiments. This holds for resting state (NODDI) and task-based (Oddball and CN-EPFL) settings. By comparing the results of models (ii) and (iv) reported in Table \ref{tab:quantitative} with the ones presented in Table \ref{tab:style}, the impact of conditioning the \textit{style} posterior vector on the attention scores is quite noticeable. And it goes beyond the simple induction of \textit{style} in the latent space, as Table \ref{tab:style} shows that placing a \textit{style} prior can cause overfitting in some settings.

\section{Discussion}\label{section:discussion}

\textbf{EEG electrode attentional based relations dependency.} The ran experiments with different types of \textit{style}, $z_w$, in the latent representation (see Equation \ref{equation:attention_style_posterior}), tell us that conditioning the styling on the attention scores, an EEG electrode topographical representation, is beneficial for the fMRI synthesis task. Further, the fact that, in addition to not conditioning \textit{style}, learning a \textit{style} prior vector is not as informative (no dependency on $\vec{x}$) for the neural network to better optimize the learning objective. This leads us to believe that a learnable unconditioned \textit{style} acting as a prior, is prone to overfitting the training data, since it is not conditioned on $\vec{x}$. Our experiments show that the projected random Fourier features (prior), $\vec{z}_x \to \vec{z}^*_x$, if multiplied (conditioned) by data dependent (EEG attention graph scores), Equation \ref{equation:attention_style_posterior}, not only reduces the empirical risk, but is also preferable to both multiplication of an unconditioned learnable \textit{style} prior and no multiplication at all. Therefore, the placement of a \textit{style} posterior, conditioned on EEG attention scores guides the random Fourier features and removes the inherent assumptions of a prior \cite{tenenbaum2011grow}. Adding to it, the topographical information retrieved from the attention scores contains information that is highly related to haemodynamical activity, this is in accordance with several neuroscience studies that use topographical structures, such as graphs, to relate EEG and fMRI, used in simultaneous EEG and fMRI studies \cite{yu2016building, rojas2018study, brechet2019capturing}.

\begin{figure*}[t]
    \centering
     \begin{subfigure}[t]{0.45\textwidth}
         \centering
         \includegraphics[clip, trim=3cm 3cm 3cm 3.5cm,width=\textwidth]{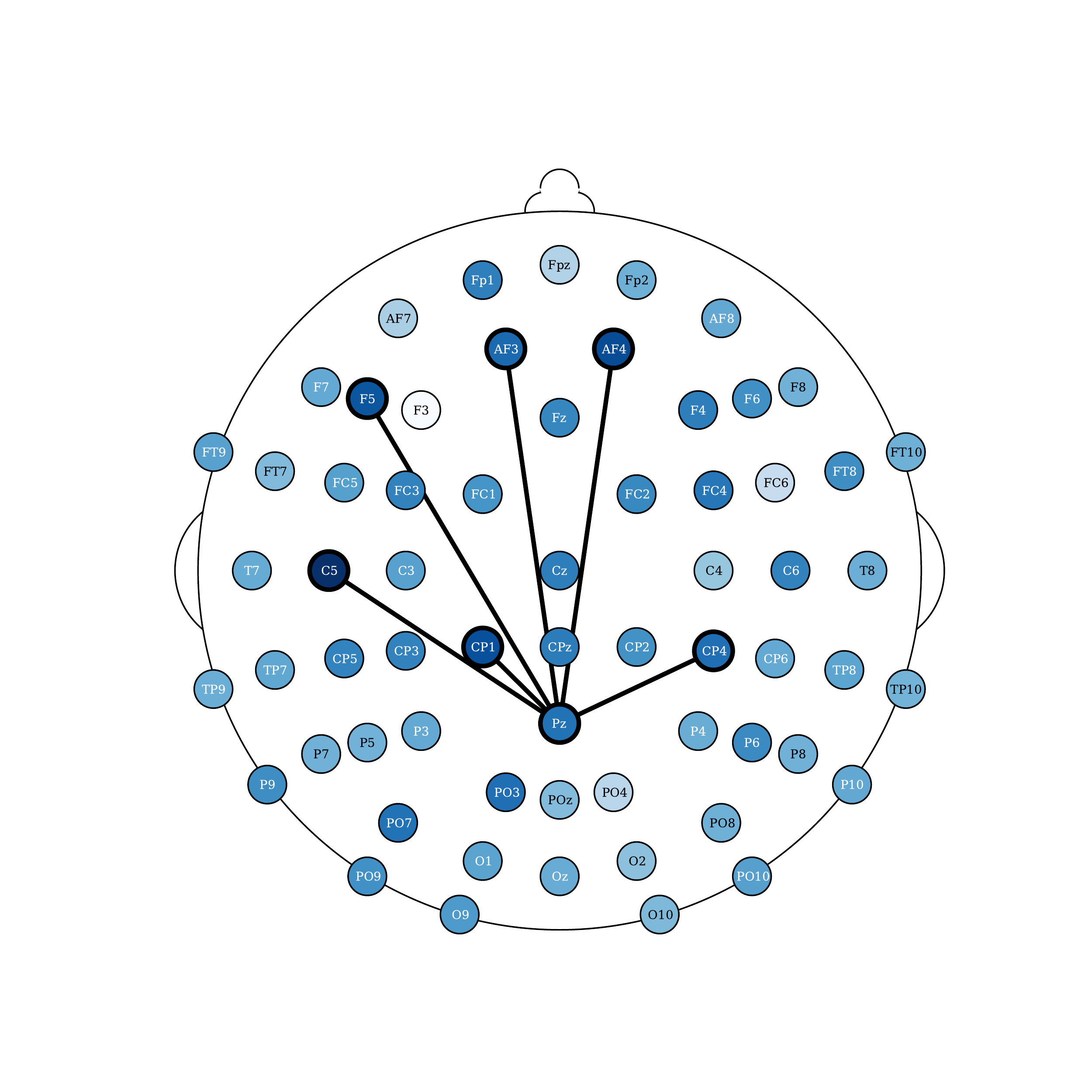}
         \caption{(iv) - Topographical attention on the EEG electrodes dimension, with random Fourier feature projections in the latent space, in NODDI dataset.}
         \label{fig:01_topo_fourier_attention_style_eeg_channels}
     \end{subfigure}
     \hfill
     \begin{subfigure}[t]{0.45\textwidth}
         \centering
         \includegraphics[clip, trim=3cm 3cm 3cm 3.5cm,width=\textwidth]{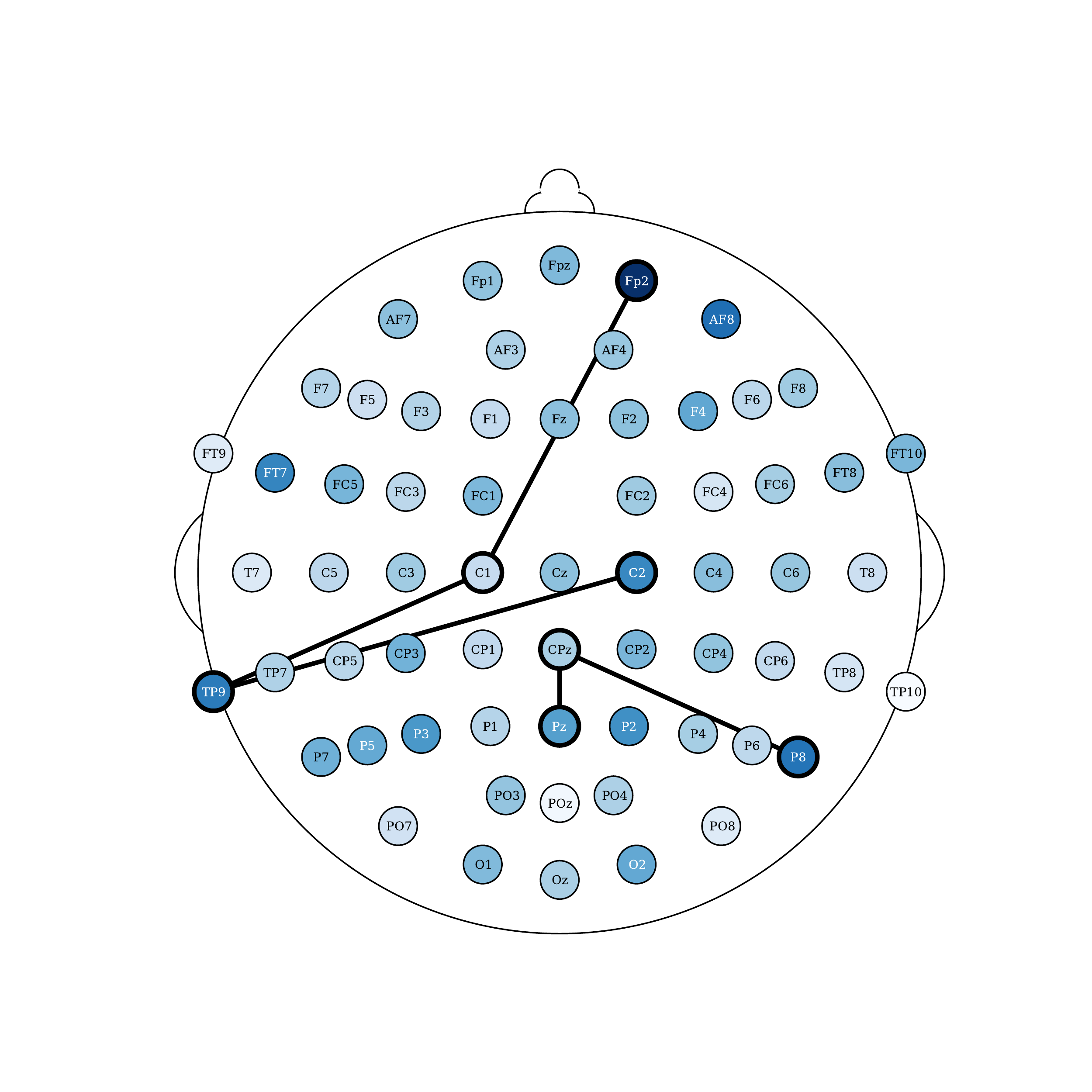}
         \caption{(iv) - Topographical attention on the EEG electrodes dimension, with random Fourier feature projections in the latent space, in CN-EPFL dataset.}
         \label{fig:03_topo_fourier_attention_style_eeg_channels}
     \end{subfigure}
     \caption{EEG electrode attention score relevances for resting state NODDI and task based CN-EPFL datasets. Figures \ref{fig:01_topo_fourier_attention_style_eeg_channels} and \ref{fig:03_topo_attention_style_eeg_channels}
     report the attention relevances for the NODDI resting state dataset and the CN-EPFL dataset, respectively.}
     \label{fig:eeg_channels_relevances}
\end{figure*}

\vskip 0.2cm\noindent\textbf{Most relevant electrode relations.} Consider the relevance of the attention scores, computed from models (ii) and (iv), both having topographical attention at the EEG channel dimension, and model (iv) with projected random Fourier features in the latent space. These relevances were propagated, using the LRP algorithm \cite{bach2015pixel} described in Appendix \ref{appendix:lrp}, through the attention style based posterior. Figure \ref{fig:eeg_channels_relevances} shows the relevances plotted in a white to blue scale, from less relevant to most relevant, respectively. The latter only shows the edges that are above the $99.7$ percentile. The presence of an edge between electrodes suggests that either this connection yields a Markovian property for the EEG instance or, otherwise, it is relevant to add fMRI style conditioned on these connections (recall from Section \ref{methods:topo_attention} that posterior $z_w$ conditions the latent EEG representation $z^*_x$ such that $z^*_x \odot z_w$). For resting state fMRI, both Figures \ref{fig:01_topo_attention_style_eeg_channels} and \ref{fig:01_topo_fourier_attention_style_eeg_channels} show connections of parietal and occipital channels (O2 electrode in Figure \ref{fig:01_topo_attention_style_eeg_channels} and Pz electrode in Figure \ref{fig:01_topo_fourier_attention_style_eeg_channels}) with frontal and central channels to be the most relevant (above the 99.7 percentile of relevance). Figure \ref{fig:01_topo_attention_style_eeg_channels} reports an additional connection between the Oz and PO9 electrodes, a correspondence between an occipital and a parietal-occipital electrode, which is in accordance with connectivity observations reported by \citet{rojas2018study}. There were no reported relevances for the electrodes (T) placed in the temporal regions for resting state settings. In contrast, in task-based fMRI synthesis, relevant relationships between temporal (FT9 and TP9) and frontal/central (Fp2 and C1/C2, respectively) electrodes were reported, see Figures \ref{fig:03_topo_attention_style_eeg_channels} and \ref{fig:03_topo_fourier_attention_style_eeg_channels}. In both of these figures, connections between central and parietal electrodes were observed. Particularly, there were reported connections between Cz with Pz and CP5 and CP2 electrodes in Figure \ref{fig:03_topo_attention_style_eeg_channels}. And connections between Pz and P8 with CPz electrodes in Figure \ref{fig:03_topo_fourier_attention_style_eeg_channels}.

\begin{figure*}[t]
    \centering
    \includegraphics[clip, trim=4cm 2cm 4cm 2cm,width=  \textwidth]{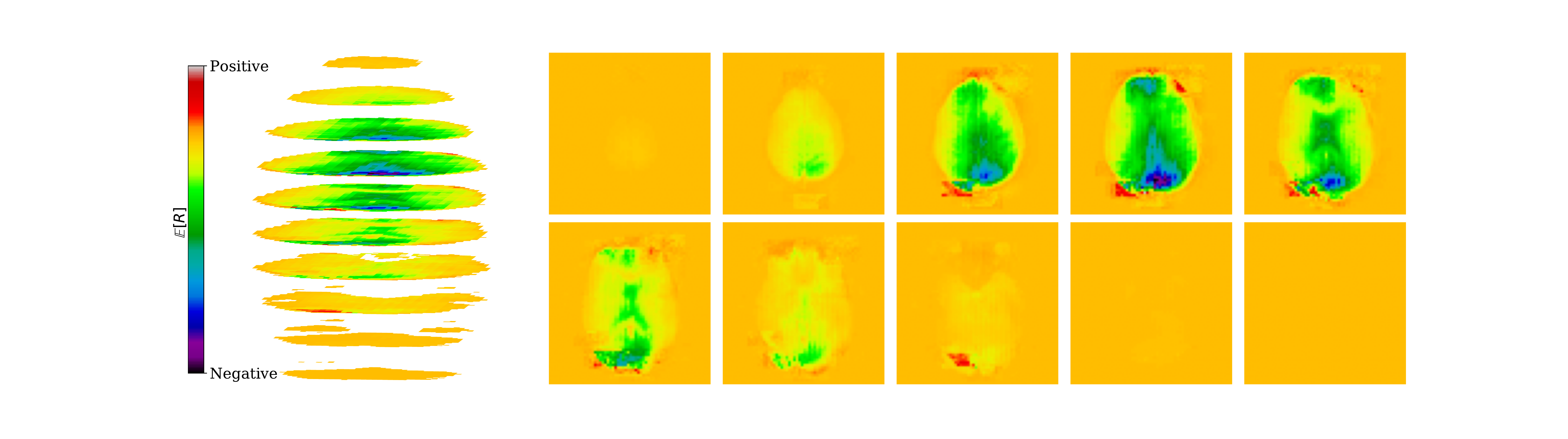}
    \caption{fMRI computed relevances for the NODDI dataset, starting from the latent fMRI representation, $\vec{z}_y$.}
    \label{fig:fmri_pvalues}
\end{figure*}

\vskip 0.25cm
\noindent\textbf{Converging to retrieve near scalp haemodynamical activity.} One interesting phenomena that was observed by propagating relevances from the latent representations of the fMRI instance, $\vec{z}_y$, to the input, $\vec{y}$, was that the relevances in sub-cortical areas were neither positive nor negative, yielding residual relevance, as seen in Figure \ref{fig:fmri_pvalues}. This later observation suggests that haemodynamical activity from these areas does not significantly aids the targeted synthesis. 
Recall that the regularization term, $\Omega(\vec{z_x}, \vec{z_y}) = cos(\vec{z_x}, \vec{z_y})$, is used with the latent EEG and fMRI representations. This is in accordance with the fact that the retrievable information is in its majority next to the scalp, where the electrodes are placed. \citet{de2019introduction} discuss how high frequencies are not able to travel significant distances with obstacles, such as white matter and the scalp, in between. We also report negative relevances on the visual cortex and positive relevances on the occipital and prefrontal lobes. Please note that negative and positive relevances represent relevant features, whereas when one has zero relevance, it means a feature was not relevant for the task. \citet{daly2019electroencephalography} found that neuronal activity retrieved from EEG can reflect the haemodynamical changes in subcortical areas. Here we claim that haemodynamical activity information in areas next to the scalp are relevant to learn the shared latent space. 

\vskip 0.25cm
\noindent\textbf{Laboratory setup impacts EEG to fMRI synthesis.} The results show that it is more difficult, according to the RMSE metric, to synthesize task-based fMRI than resting state. This observation is in contrast with studies that report that resting state fMRI is inherently more complex than task based fMRI \cite{maknojia2019resting}. The SSIM metric, in contrast to the RMSE, shows less significant differences for the Oddball recordings in favor of fMRI synthesis in the resting state. 
However, the CN-EPFL dataset is not in accordance with the latter. This performance heterogeneity across the datasets may not only rise from the characteristics of the recording sessions, but may also be also propelled by the different preprocessing techniques employed. Each dataset is publicly available and is supported with published studies, having unique equipment, experimental protocols, and algorithms. CN-EPFL dataset is the most complete one, with a total of 20 individuals and with a resolution of $2 \times 2 \times 2$mm, which makes a total of $108 \times 108 \times 64$ voxels. These differences, caused by working with 3 Tesla (CN-EPFL dataset) versus 1.5 Tesla (NODDI and Oddball datasets) scanners, significantly impact the spatial resolution, which for the datasets NODDI and Oddball produce $64 \times 64 \times 30$ and $64 \times 64 \times 32$ voxels, respectively, with around $3\times 3 \times 3$mm voxel size. 
One has to further account for the original recording artifacts and disruptions caused by the applied preprocessing techniques. For instance, Oddball dataset contains intra and inter individual wise misalignments across fMRI volumes. This may be the cause of poorer performance of all methods when compared to the other datasets. 
In addition, 
Oddball relies on a different EEG electrode positioning system, having a total of $43$ electodes that were not placed in accordance with the 10-20 system \cite{jasper1958ten}. Although NODDI and CN-EPFL recordings are in accordance with this system, each selected unique electrode locations (see the different electrode placements between Figures \ref{fig:01_topo_fourier_attention_style_eeg_channels} and \ref{fig:03_topo_fourier_attention_style_eeg_channels}). Finally, the different EEG sampling frequencies, with $250$Hz, $1000$Hz and $5000$Hz considered in NODDI, Oddball and CN-EPFL recordings, respectively, further affect architectural operations and subsequently impact the learning. 

\section{Related work}\label{section:related_work}

\textbf{EEG and fMRI synthesis research state.} The learning of mapping functions between structural neuroimaging modalities is increasingly prevalent, e.g. transfer functions between MR and CT scans achieved arguable success using convolutional networks \cite{dong2017, wolterink2017}. For a comprehensive description of multi modal brain structural image synthesis, please refer to \cite{YI2019101552}. In contrast, regression between functional neuroimaging modalities has not received the same amount of attention. The pivotal work by \citet{liu2019convolutional} relies on diverse convolutional operations to pursue the fMRI synthesis from EEG. The work in its entirety goes further and performs bi-directional synthesis. Nonetheless, and despite the efforts made by \citet{liu2019convolutional}, advances in this field remain to be explored. 
Related studies relate haemodynamics with a second (non-neurological) modality using optimized transformations \cite{cury2020sparse, raposo2022learning}, while others work directly in the regression of localized haemodynamical activity in the context of natural language \citet{jain2021interpretable} or music retrieval tasks \citet{hoefle2018identifying}. All of these works show the feasibility of haemodynamical retrieval and its usefulness to enrich information. The lack of exploration of cross modal functional neuroimaging arises from various factors inherently present in the structural and representational dissimilarities between modalities. 
In contrast with structural neuroimaging techniques, the alignment of functional neuroimaging modalities in time has to be further ensured. 
Simultaneous EEG and fMRI recordings \cite{dataset_01, dataset_02, dataset_03} 
have been increasing conducted, representing a notable effort by the research community that can pave the way to further breakthroughs on the fMRI synthesis from EEG signals. 
\vskip 0.25cm

\noindent\textbf{Simultaneous EEG and fMRI studies.} \citet{chang2013eeg} claim decreases in alpha band and increases in theta band in the time dimension are correlated with relative increases in functional connectivity.
\citet{cury2020sparse} combined EEG and fMRI to build an EEG informed fMRI modality, that is cheaper than fMRI and contains informed features.
\citet{he2018spatialtemporaldo} report positive correlation between the haemodynamical activity with alpha band, showing that the temporal resolution of spectral information is important to address the combination of these modalities.
\citet{leite2013transfer} associated EEG spectral features with haemodynamicall activity for a single epileptic subject.
Similarly, \citet{rosa2010estimating} observe that changes in haemodynamical activity were also in accordance with changes in spectral features of neuronal activity. Results gathered in the context of our work further confirm the majority of the aforementioned findings.


\section{Conclusion}\label{section:conclusion}

We found that topographical relationships between EEG channels are highly relevant and beneficial for the targeted fMRI synthesis task. Our experiments conclude that attention-based scores, trained to give Markovian properties to the EEG representation and simultaneously add style features by usage of a posterior, significantly aid the task.
Relationships learned between occipital, parietal and frontal electrodes were observed to be of particular relevance to retrieve haemodynamical activity. We further noticed that haemodynamical information in areas next to the scalp is predominantly considered to learn the shared latent space during the training, aiding fMRI synthesis.

We hope to have motivated researchers to work in this emerging field. 
Neuroimaging synthesis and augmentation from more accessible modalities yields unique opportunities for reducing costs and improving diagnostics in health care settings, while offering ambulatory and longitudinal proxy views of haemodynamical activity. Future work is expected to validate the feasibility of the fMRI synthesis task for the aforementioned ends, comprehensively assessing the predictive limits of electrophysiological activity measured at the cortex.

\section{Acknowledgments}\label{section:acknowledgments}

This work was supported by national funds through Funda\c{c}\~ao para a Ci\^encia e Tecnologia (FCT), under the Ph.D. Grant SFRH/BD/5762/2020 to David Calhas, ILU project DSAIPA/DS/0111/2018 and INESC-ID pluriannual UIDB/50021/2020. We thank Alexandre Francisco and Sérgio Pereira for their very useful input to the work, in the context of the PhD advising committee. We want to give a special thanks to João Rico, Daniel Gonçalves, Pedro Orvalho and Leonardo Alexandre for giving amazing feedback on the visual support used in this study. We also want to thank António Gusmão for the enriching discussions had on the role of attention mechanisms used.

{\small
\bibliographystyle{unsrtnat}
\bibliography{references}
}

\newpage
\appendix

\section{Automatic generation of neural architectures}\label{setting:auto_nas}

\citet{calhas2022automatic} proposed a neural architecture generation method that respects the arithmetic of convolutions with a \textit{valid} padding. The relation between the input, $I$, and output, $O$, of a downsampling layer (e.g., convolution, pooling) is in accordance with 

\begin{equation}\label{equation:convolution}
    O = \frac{I-k}{s} + 1 \Leftrightarrow I = (O-1)\times s + k,
\end{equation}

where $k$ and $s$ are the kernel and stride sizes, respectively. As such, a neural network specification composed of downsampling layers is formalized as 

\begin{equation}\label{equation:na_specification}
    \forall n \in \{1, \dots, N\}: L_n = (k_n, s_n),    
\end{equation}

being $L$ the notation for a layer. In terms of layer processing, the function of a layer is $f_{L_n}: \mathbb{R}^{I_n} \mapsto \mathbb{R}^{O_n}$, where $I_n$ is the input of the $n$th layer and $O_n$ the output. Similarly, the function that represents the neural network $f: \mathbb{R}^I \mapsto \mathbb{R}^O \wedge f(x) = f_{L_N}(f_{L_{N-1}}(\dots f_{L_1}(\vec{x})\dots))$. Using Satisfiable Modulo Theory and with the encoding

\begin{equation}\label{equation:auto_nas}
    O = O_N \wedge \bigwedge_{n=1}^N O_n \leq O_{n-1} \wedge k_n > 0 \wedge s_n > 0,
\end{equation}

one can use a solver \cite{z3_citation} to get an assignment to the kernel, $k$, and stride, $s$, of all layers. The formulation is further extended to multiple dimensions and variable number of layers. For details on the latter please refer to \cite{calhas2022automatic}. This approach is used in order to remove the human bias from the methodology proposed.

\begin{figure}[t]
    \centering
    \includegraphics[width=0.5\textwidth]{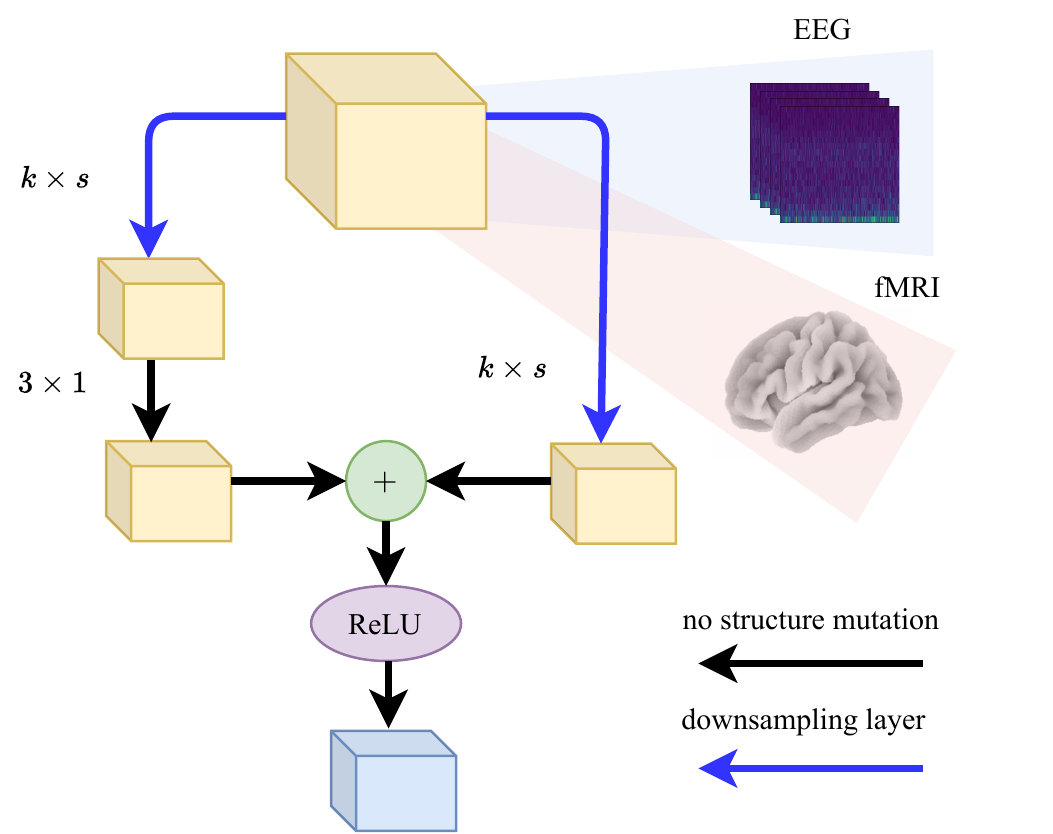}
    \caption{The inspired Resnet-18 block forks the input in two computational flows: (1) the first, represented in the left part of the figure, is processed by a convolutional layer with $k \times s$ as the kernel and stride sizes operate with \textit{valid} padding, following the output goes through a convolutional layer with $3 \times 1$ with \textit{same} padding; (2) the second flow, corresponds to the right arrow of the fork, processes the input with a convolutional layer with $k \times s$ with a \textit{valid} padding. The representations of the fork are joined by the \textit{addition} operation, which is followed by a ReLU activation \cite{nair2010rectified}. Please note that max pooling \cite{nagi2011max} and batch normalization \cite{ioffe2015batch} layers are optional to follow each downsampling layer. EEG and fMRI feature representations are included in the figure for the reader to understand that this block structure is used to process EEG and fMRI, though differing in the values of $k \times s$ in each network.}
    \label{fig:resnet_conf}
\end{figure}

\textbf{Resnet-18 block configuration. } The neural architecture specification (Equation \ref{equation:na_specification}) is to replace the downsampling blocks of the Resnet-18, illustrated in Figure \ref{fig:resnet_conf}. \citet{he2016deep} defined kernel and stride sizes set as $1 \times 2$ for all blocks. Note that, it is encouraged to use different kernel and stride sizes for the different layers \cite{anonymous2022learning}. Therefore, the variable assignments, of Equation \ref{equation:na_specification}, are used as the kernel and stride sizes.

\section{Hyperparameters and latent dimension}\label{results:latent}

Table \ref{tab:hyperparameters} reports on the hyperparameters obtained from running the Bayesian optimization algorithm, with the setup described in Section \ref{setting:bayesian_optimization}.

\begin{table*}[t]
    \centering
    \scriptsize{
    \begin{tabular}{p{1.2cm} | p{1.2cm} | p{1.2cm} | p{1.2cm} | p{1.2cm} | p{1.2cm} | p{1.2cm} | p{1.2cm} | p{1.2cm}}
        \hline
        \hline
        \hfil lr & \hfil $\theta$ decay & \hfil batch size & \hfil $K$ & \hfil filters & \hfil max pool & \hfil batch norm & \hfil skip & \hfil dropout \\
        \hline
        \hfil $2.98e-3$ & \hfil $4.40e-4$ & \hfil $4$ & \hfil $7$ & \hfil $4$ & \hfil $1$ & \hfil $1$ & \hfil $1$ & \hfil $0.5$  \\
        \hline
        \hline
    \end{tabular}
    \caption{Hyperparameters obtained from the Bayesian optimization algorithm, ran for $100$ iterations.}
    \label{tab:hyperparameters}
    }
\end{table*}

\begin{figure}[ht]
    \centering
    \includegraphics[width=0.45\textwidth]{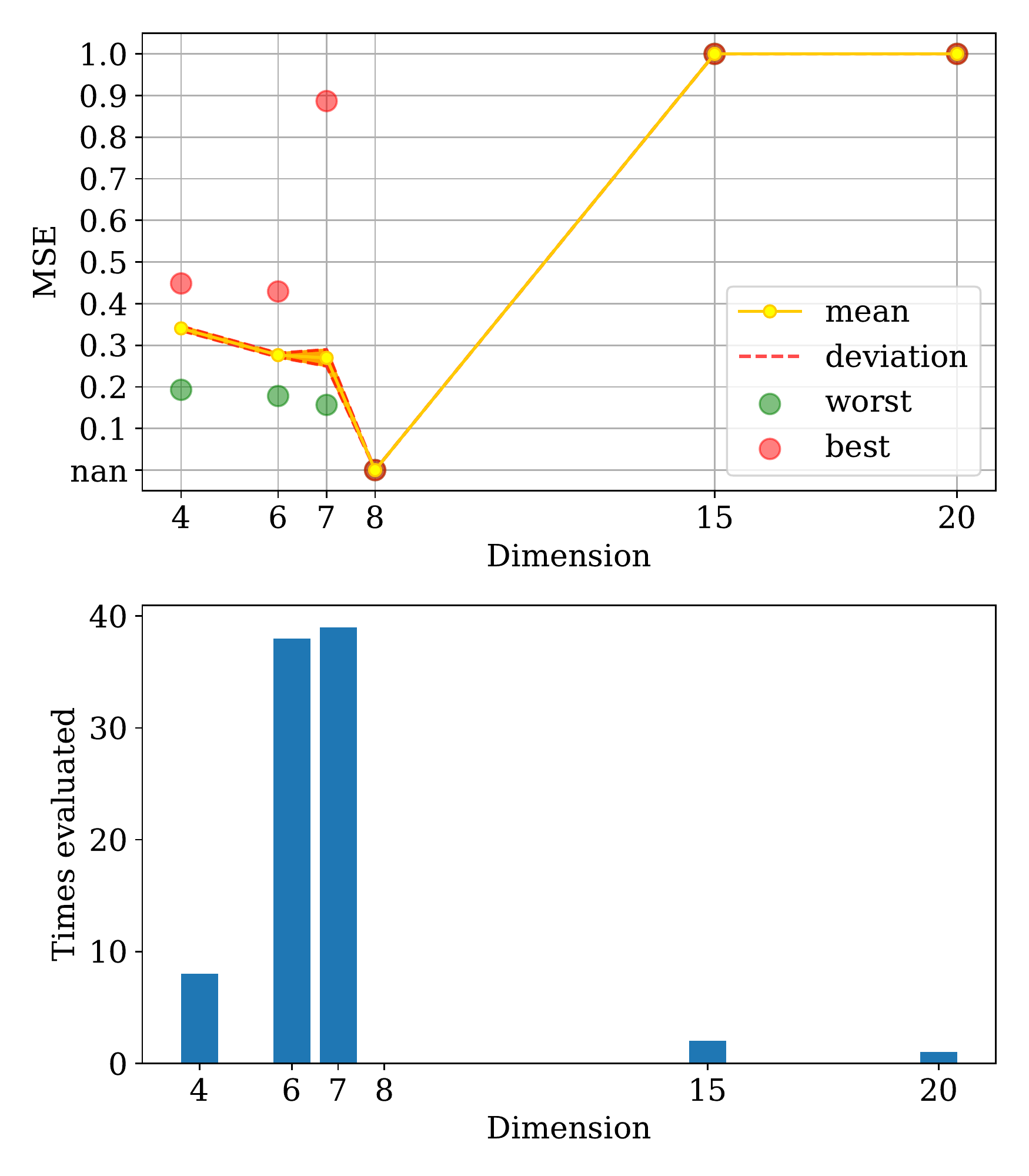}
    \caption{Latent dimension analysis according to the MSE metric.}
    \label{fig:latent_analysis}
\end{figure}

Figure \ref{fig:latent_analysis} shows the performance of each evaluation made during the hyperparameter search. $K=7$ achieved the best result, followed by $K=6$ and $K=4$. Interestingly, $K=8$ produced non defined values due to the training being underway, but GPU memory was exceeded. Consequently, these were not counted as evaluations. As for $K=15$ and $K=20$, the model was not able to be loaded to the GPU and failed to be evaluated. In contrast with $K=8$, $K=15$ and $K=20$ were evaluated since it did not freeze the GPU. Memory limitation is important, because the target of this framework is to be achievable in a day-to-day laptop, enabling the use of this work in a cheap setup.

\section{Neural networks generated}\label{results:candidates}

\begin{figure*}[t]
    \centering
    \begin{subfigure}[t]{0.48\textwidth}
        \centering
        \includegraphics[width=\textwidth]{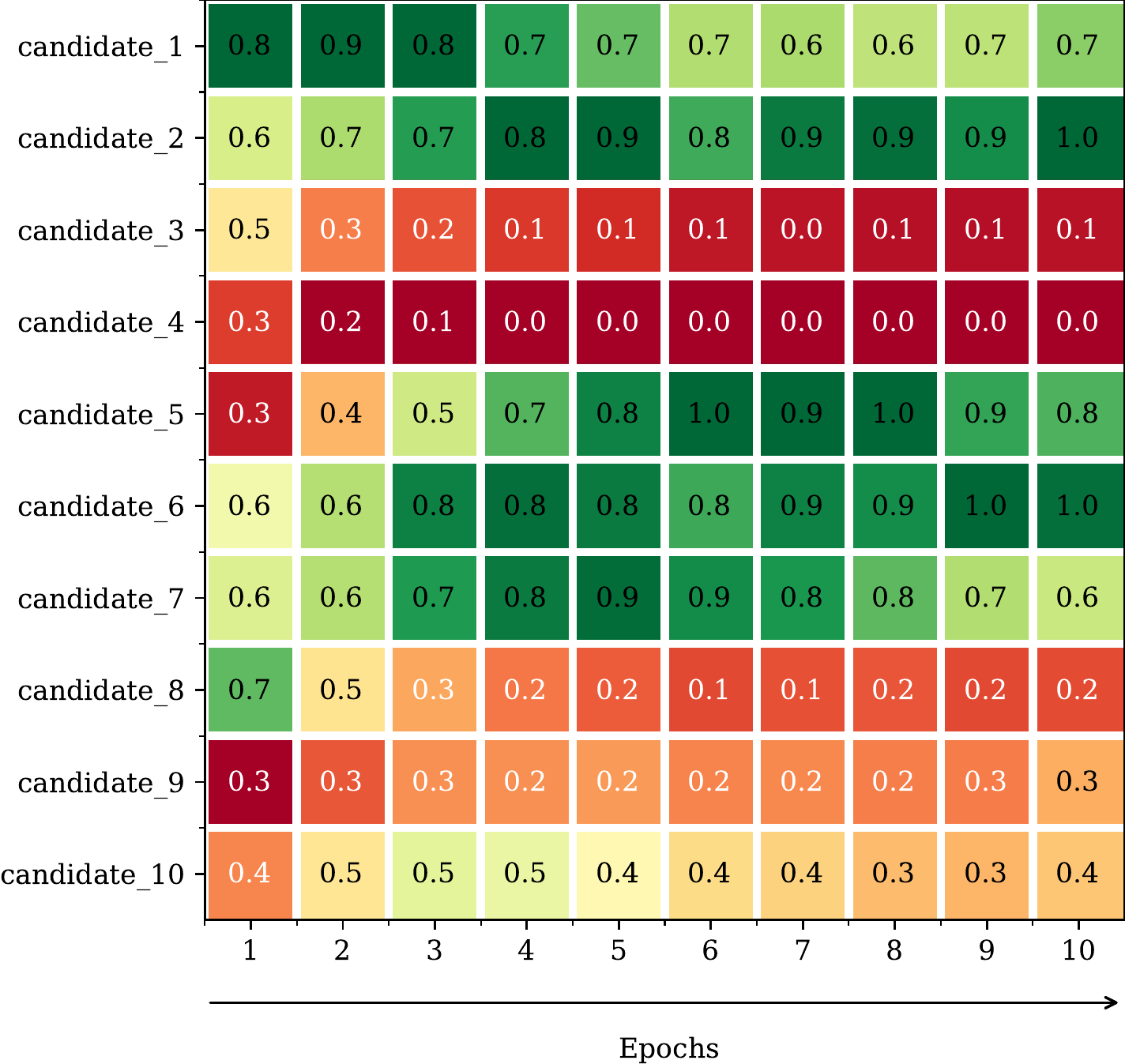}
        \caption{DARTS weight deviation for the generated NAs for the EEG encoder.}
        \label{fig:eeg_darts_convergence}
    \end{subfigure}
    \hfill
    \begin{subfigure}[t]{0.48\textwidth}
        \centering
        \includegraphics[width=\textwidth]{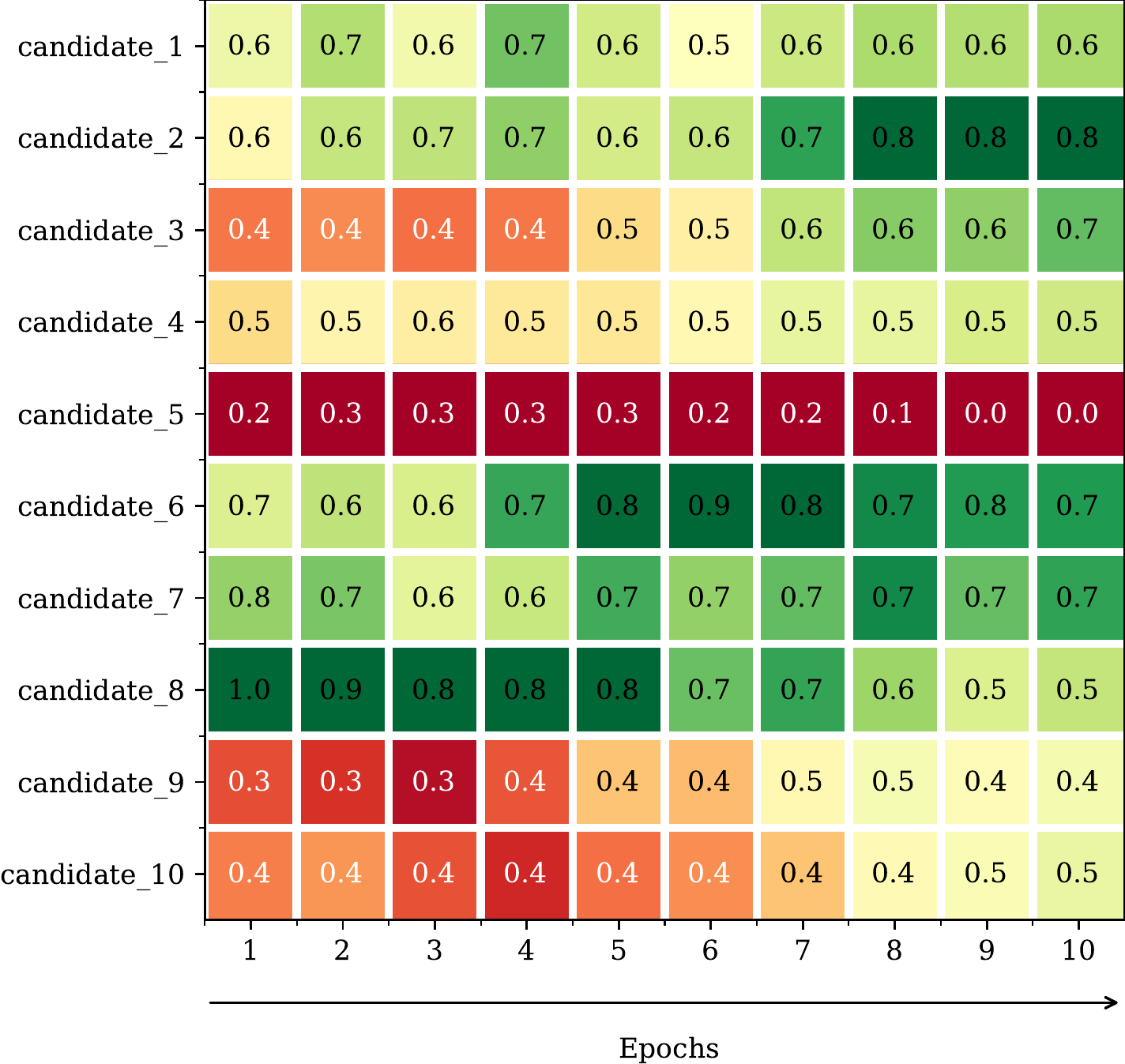}
        \caption{DARTS weight deviation for the generated NAs for the fMRI encoder.}
        \label{fig:fmri_darts_convergence}
    \end{subfigure}
    \caption{\citet{liu2018darts} algorithm weight convergence for each network, for the EEG and fMRI encoder.}
    \label{fig:darts_convergence}
\end{figure*}

In this Section, the generated neural architectures, using the method described in Section \ref{setting:auto_nas} and with more detail in \cite{calhas2022automatic}, are reported. The neural architecture search was performed with the 01 dataset. The latter, means that the EEG instance $\vec{x} \in \mathbb{R}^{64 \times 134 \times 10}$ and the fMRI instance $\vec{y} \in \mathbb{R}^{64 \times 64 \times 30}$. This search took a total of 3 months, due to the limited GPU resources, as well as the memory of the GPU that was used. In practice, this computational time can be reduced if one does not train all the generated networks at the same time, which means \cite{liu2018darts} algorithm would not be used. Despite of it, using the algorithm provides information about the convergence of each network simultaneously, as shown in Figures \ref{fig:eeg_darts_convergence} and \ref{fig:fmri_darts_convergence}. 

\textbf{EEG candidates.} Starting with the EEG encoder, which was generated setting $I=64 \times 134 \times 10$ and $O=7 \times 7 \times 7$, the properties, such as the kernel, stride and number of layers, of the architecture are presented in Table \ref{tab:eeg_candidates}. Along with the properties of the architectures generated, the convergence of the \cite{liu2018darts} algorithm is reported in Figure \ref{fig:eeg_darts_convergence}. By analyzing the Figure, we conclude that the best architecture obtained was the candidate number 2.

\begin{table*}[t]
    \centering
    \begin{tabular}{p{1.5cm} | p{7.5cm} | p{1.5cm}}
    
        \hline
        \hline
         \hfil Candidate & \hfil Kernel $\times$ Stride ($\bigwedge_{1}^N k^{(1)},k^{(2)},k^{(3)} \times s^{(1)},s^{(2)},s^{(3)}$) & \hfil $N$\\
         \hline
         \hfil 1 & \hfil $11,86,2 \times 1,1,1 \wedge 17,20,2\times 4,2,1 \wedge 2,7,2\times 1,1,1$ & \hfil 3\\
         \hfil 2 & \hfil $7,37,2\times 3,5,1 \wedge 7,7,2\times 2,2,1$  & \hfil 2\\
         \hfil 3 & \hfil $9,43,2\times 1,2,1 \wedge 11,11,2\times 1,2,1 \wedge 9,3,2 \times 5,2,1$  & \hfil 3\\
         \hfil 4 & \hfil $28,15,2\times 1,1,1 \wedge 30,77,2\times 1,7,1$  & \hfil 2\\
         \hfil 5 & \hfil $7,19,2\times1,1,1 \wedge 20,23,2\times 1,4,1 \wedge 23,16,2 \times 2,1,1$  & \hfil 3\\
         \hfil 6 & \hfil $6,29,2\times1,1,1 \wedge 21,33,2\times1,4,1 \wedge 16,11,2\times 3,1,1$  & \hfil 3\\
         \hfil 7 & \hfil $32,47,2\times 2,4,1 \wedge 4,15,2 \times 2,1,1$  & \hfil 2\\
         \hfil 8 & \hfil $9,16,2\times 3,1,1 \wedge 5,2,2\times 1,1,1 \wedge 6,81,2 \times 1,5,1$  & \hfil 3\\
         \hfil 9 & \hfil $23,32,2\times 1,1,1 \wedge 11,96,2 \times 5,1,1$  & \hfil 2\\
         \hfil 10 & \hfil $16,31,2\times 1,8,1 \wedge 24,6,2\times 4,1,1$  & \hfil 2\\
         \hline
         \hline
    \end{tabular}
    \caption{From input shape $64 \times 134\times10 $ to output shape $K\times K \times K$ with $K=7$. Each layer is followed by a max-pool operation with $2,2,1\times 1,1,1$.}
    \label{tab:eeg_candidates}
\end{table*}

\textbf{fMRI candidates.} Following with the fMRI encoder, which was generated setting $I=64 \times 64 \times 30$ and $O=7 \times 7 \times 7$, the properties of the architecture are presented in Table \ref{tab:fmri_candidates}. By analyzing Figure \ref{fig:fmri_darts_convergence}, we conclude that the best architecture obtained was the candidate number 2.

\begin{table*}[t]
    \centering
    \begin{tabular}{p{1.5cm} | p{7.5cm} | p{1.5cm}}
        \hline
        \hline
         \hfil Candidate & \hfil Kernel $\times$ Stride ($\bigwedge_{1}^N k^{(1)},k^{(2)},k^{(3)} \times s^{(1)},s^{(2)},s^{(3)}$) & \hfil $N$\\
         \hline
         \hfil 1 & \hfil $16,8,8\times 4,2,1 \wedge 2,16,9\times 1,1,1 \wedge 3,5,6 \times 1,1,1$ & \hfil 3\\
         \hfil 2 & \hfil $16,6,12\times 2,1,1 \wedge 6,4,6\times 1,1,1 \wedge 12,47,5\times 1,1,1$ & \hfil 3\\
         \hfil 3 & \hfil $8,15,3\times 1,4,1 \wedge 38,6,21 \times 3,1,1$ & \hfil 2\\
         \hfil 4 & \hfil $8,7,15\times 1,1,1 \wedge 20,5,2\times 1,1,1 \wedge 15,10,6\times 3,6,1$ & \hfil 3\\
         \hfil 5 & \hfil $6,20,2\times 5,1,1 \wedge 5,8,16 \times 1,6,2$ & \hfil 2\\
         \hfil 6 & \hfil $6,44,15\times 1,1,1 \wedge 28,7,5\times 1,1,1 \wedge 16,6,3 \times 2,1,1$ & \hfil 3\\
         \hfil 7 & \hfil $14,13,5\times 1,1,2 \wedge 18,16,2 \times 1,1,1 \wedge 11,21,3 \times 3,2,1$ & \hfil 3\\
         \hfil 8 & \hfil $8,11,14\times 1,1,1 \wedge 29,19,6 \times 1,1,1 \wedge 20,27,3 \times 1,1,1$ & \hfil 3\\
         \hfil 9 & \hfil $7,2,7 \times 1,1,1 \wedge 29,25,9 \times 1,1,1 \wedge 21,23,7 \times 1,2,1$ & \hfil 3\\
         \hfil 10 & \hfil $17,28,5\times 1,1,1 \wedge 19,16,7 \times 1,1,1 \wedge 7,6,4\times 3,2,2$ & \hfil 3\\
         \hline
         \hline
    \end{tabular}
    \caption{From input shape $64 \times 64\times 30 $ to output shape $K\times K \times K$ with $K=7$. Each layer is followed by a max-pool operation with $2,2,2\times 1,1,1$.}
    \label{tab:fmri_candidates}
\end{table*}

\section{Neural architectures generation setup}\label{setting:generated_nas}

\begin{table}[t]
    \centering
    \begin{tabular}{ p{1.8cm} | p{2cm} p{2cm} }
        \hline
        \hline
        \hfil Variable & \hfil $E_x$ & \hfil $E_y$\\
        \hline
        \hfil $I$ & \hfil $64 \times 134 \times 10$ & \hfil $64 \times 64 \times 30$\\
        \hfil $O$ & \hfil $K \times K \times K$ & \hfil $K \times K \times K$\\
        \hfil $n$ & \hfil $1$ & \hfil $1$\\
        \hfil $N$ & \hfil $3$ & \hfil $5$\\
        \hline
        \hline
    \end{tabular}
    \caption{Variable specification for the formula in Equation \ref{equation:auto_nas}. The $I$ for $E_x$ and $E_y$ correspond to the EEG and fMRI representations, respectively, obtained from the \cite{dataset_01} dataset.}
    \label{tab:generated_nas}
\end{table}

Table \ref{setting:generated_nas} contains the specification of the variables to generate the neural architectures reported in Tables \ref{tab:eeg_candidates} and \ref{tab:fmri_candidates}. Following, the DARTS algorithm \cite{liu2018darts} was ran on the generated NA search space, for both $E_x$ and $E_y$, with a total of $10$ epochs, learning rate of $0.001$. Note that gradients are passed according to zero order, i.e., the weights of each network are trained independently from the weights of the softmax final layer, proposed in \cite{liu2018darts}.\footnote{The computational resources to compute all the gradients would require large amounts of GPU allocated memory, that surpasses the used NVIDIA GeForce RTX 208 GPU (8GB) capacity.} Similar to the hyperparameter optimization, a total of $8$ individuals were considered and a split of $75/25$ was done to define the training and validation sets.

\section{Layer-wise relevance propagation}\label{appendix:lrp}

\citet{bach2015pixel} proposed a method to propagate relevances from the output of a neural network to the input features. This provides relevance features, that have an informative explainability nature, assessing which ones were more relevant (either negatively or positively). Let $j$ be a hidden neuron, following the proposed propagation rule, then its relevance is computed as

\begin{equation}\label{equation:lrp}
    R_j = \sum_k \frac{a_jw_{jk}}{\sum_j a_jw_{jk}} R_k,
\end{equation}

where a neuron, $k$, has a relevance, $R_k$, associated to it. The relevance of all the neurons of the output layer are by default the output logits and the relevance of all layers are computed by backpropagation of relevances using the rule stated in Equation \ref{equation:lrp}. Note that, this rule does not apply to propagate through sinusoidal activations, which are used in this work (see Section \ref{methods:fourier}). For EEG features, the relevances are propagated through the proposed \textit{style} posterior (see Section \ref{methods:topo_attention}), where standard layers, that enable the use of this rule are used.

\begin{figure*}[t]
    \centering
     \begin{subfigure}[t]{0.45\textwidth}
         \centering
         \includegraphics[clip, trim=3cm 3cm 3cm 3.5cm,width=\textwidth]{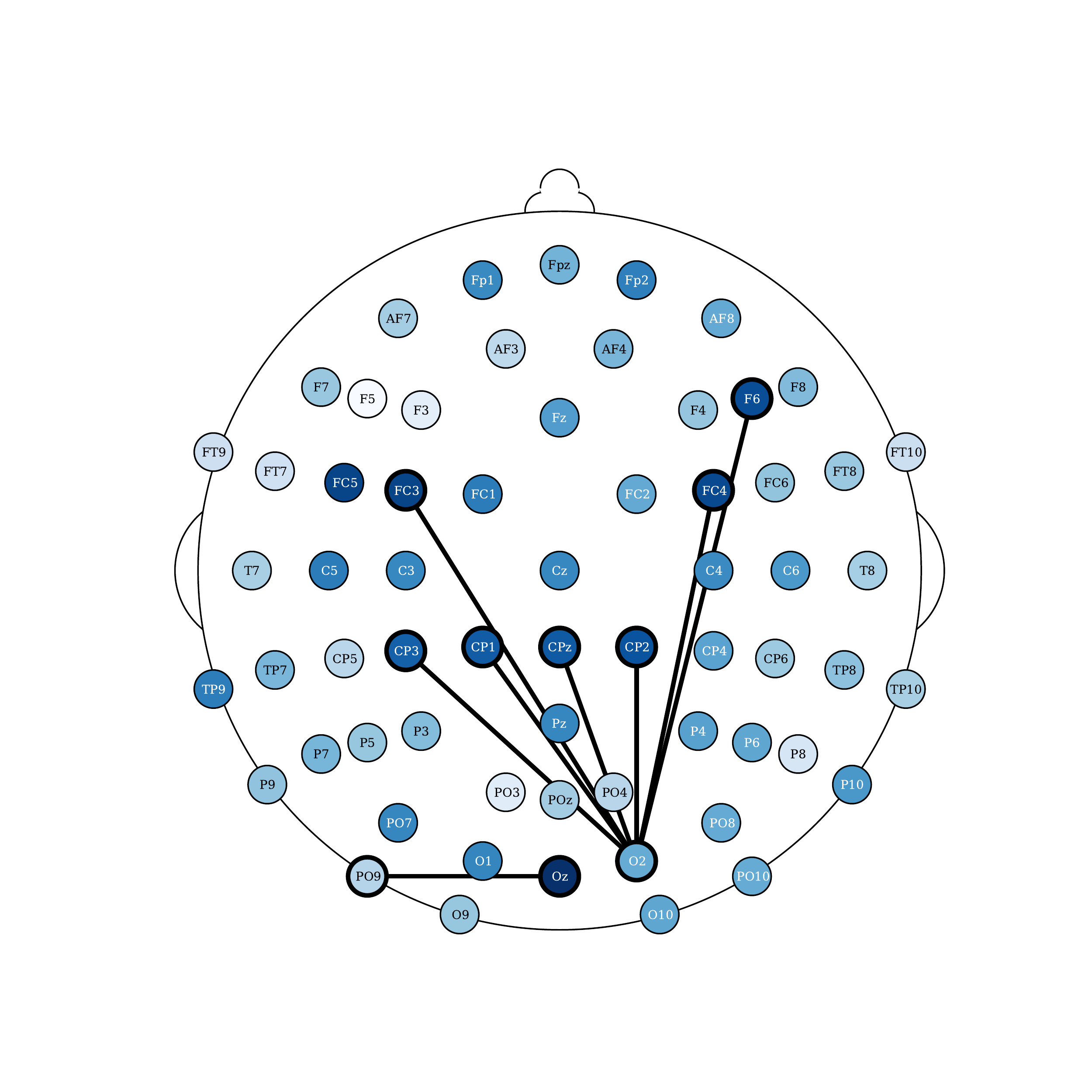}
         \caption{(ii) - Topographical attention on the EEG electrodes dimension in NODDI dataset.}
         \label{fig:01_topo_attention_style_eeg_channels}
     \end{subfigure}
     \hfill
     \begin{subfigure}[t]{0.45\textwidth}
         \centering
         \includegraphics[clip, trim=3cm 3cm 3cm 3.5cm,width=\textwidth]{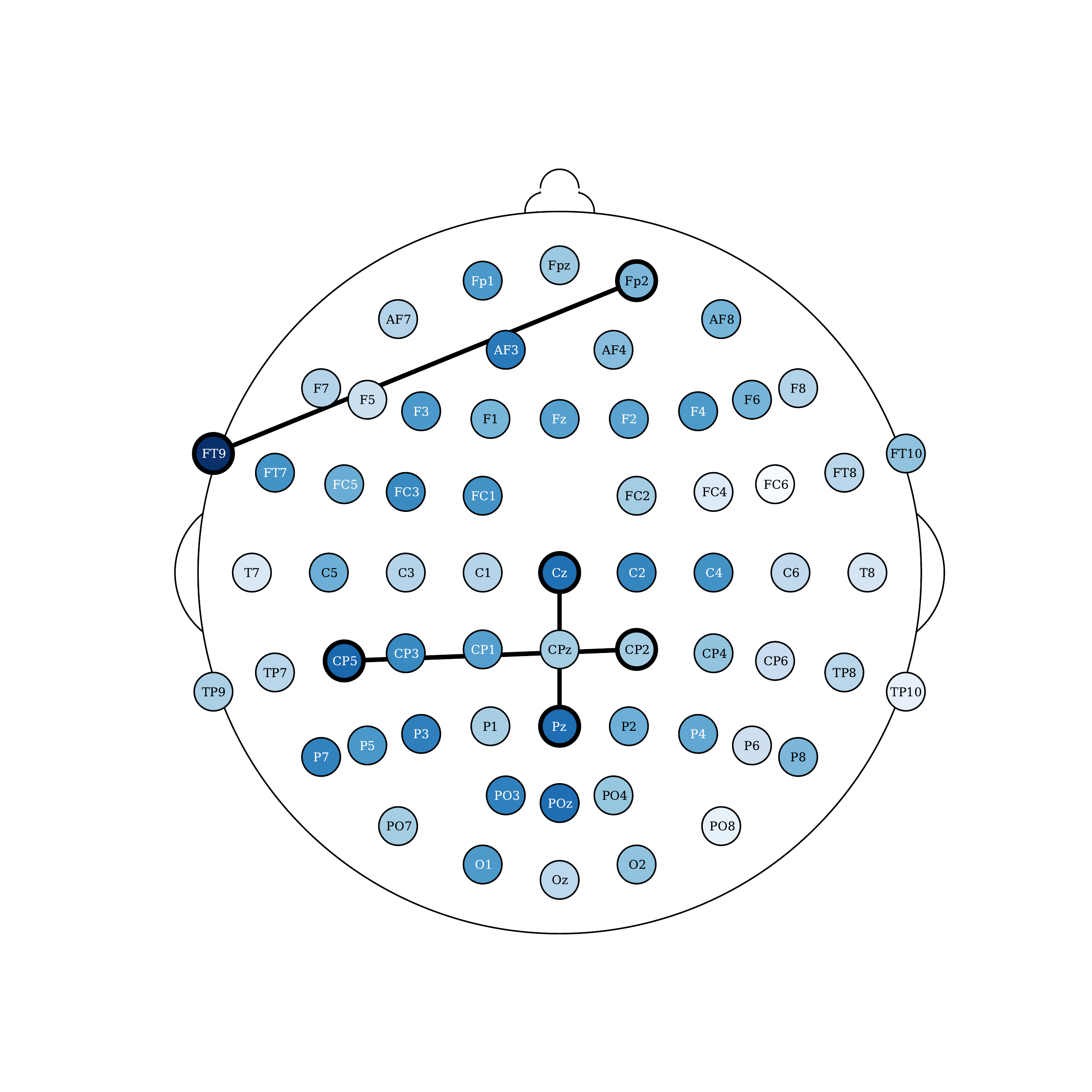}
         \caption{(ii) - Topographical attention on the EEG electrodes dimension in CN-EPFL dataset.}
         \label{fig:03_topo_attention_style_eeg_channels}
     \end{subfigure}
     \caption{EEG electrode attention score relevances for resting state NODDI and task based CN-EPFL datasets. Figures \ref{fig:01_topo_attention_style_eeg_channels} and \ref{fig:03_topo_attention_style_eeg_channels} report the attention relevances for the NODDI resting state dataset and the CN-EPFL dataset, respectively.}
     \label{fig:ii_eeg_channels_relevances}
\end{figure*}

\begin{figure*}[t]
    \centering
    \begin{subfigure}[t]{0.63\textwidth}
         \centering
         \includegraphics[clip, trim=5cm 9cm 5cm 13.5cm,width=\textwidth]{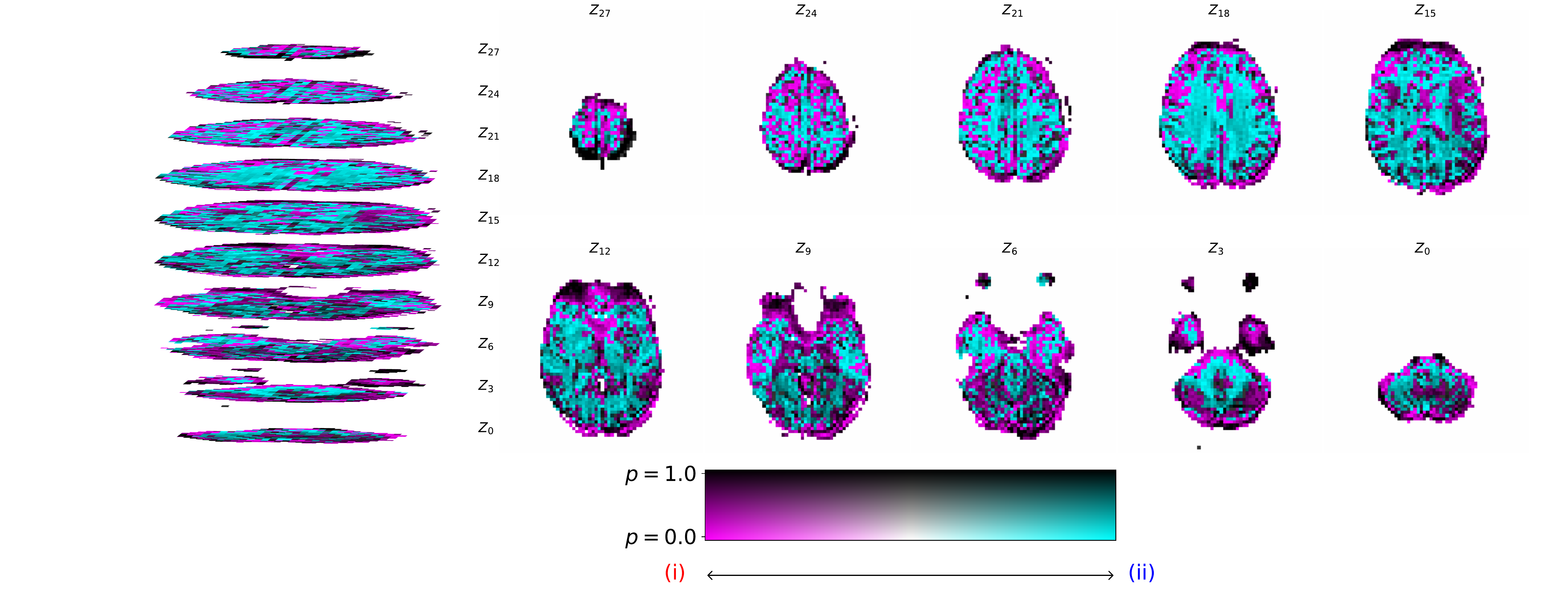}
     \end{subfigure}
     ~
     \begin{subfigure}[t]{0.63\textwidth}
         \centering
         \includegraphics[clip, trim=5cm 9cm 5cm 13.5cm,width=\textwidth]{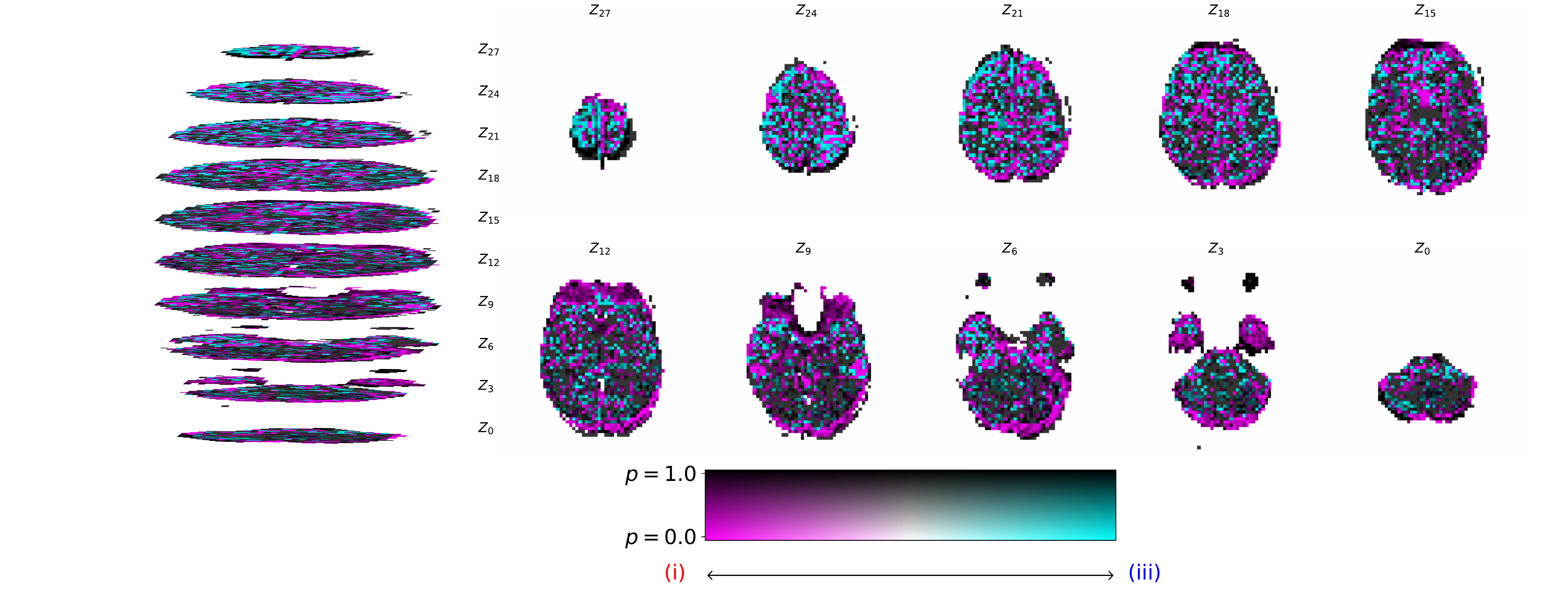}
     \end{subfigure}
     ~
     \begin{subfigure}[t]{0.63\textwidth}
         \centering
         \includegraphics[clip, trim=5cm 9cm 5cm 13.5cm,width=\textwidth]{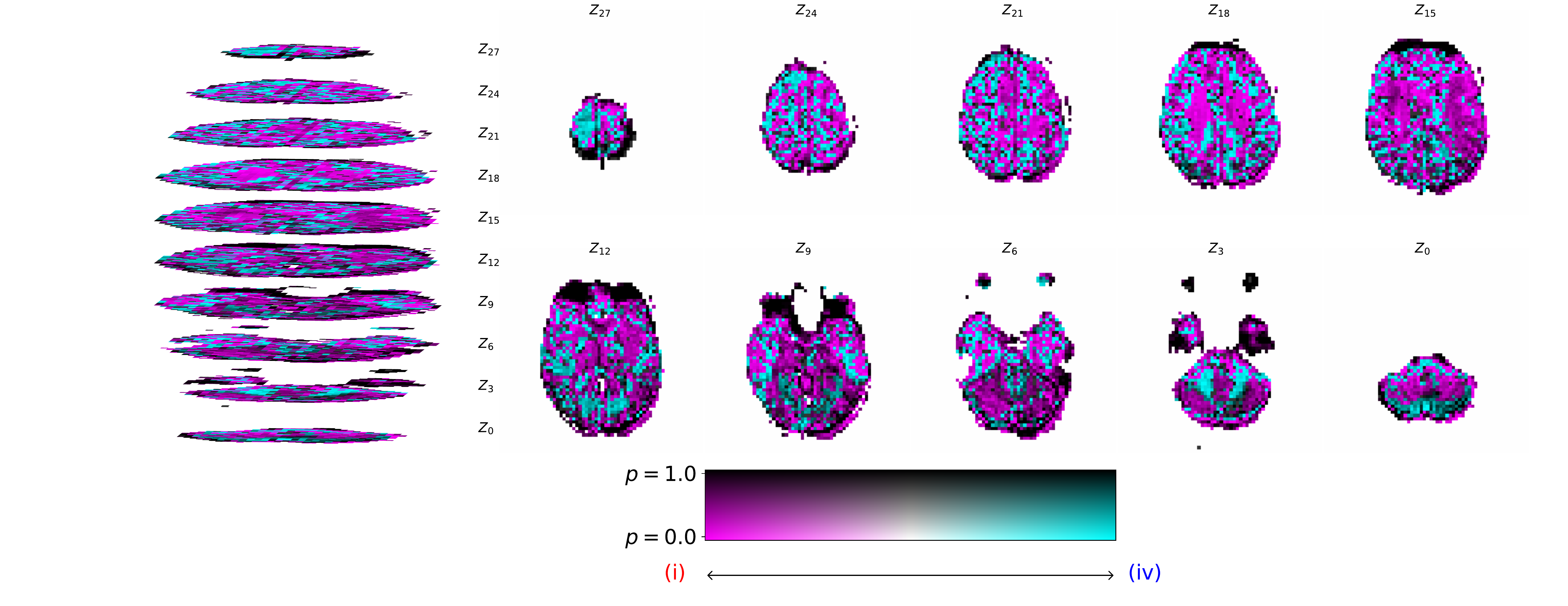}
     \end{subfigure}
     ~
     \begin{subfigure}[t]{0.63\textwidth}
         \centering
         \includegraphics[clip, trim=5cm 9cm 5cm 13.5cm,width=\textwidth]{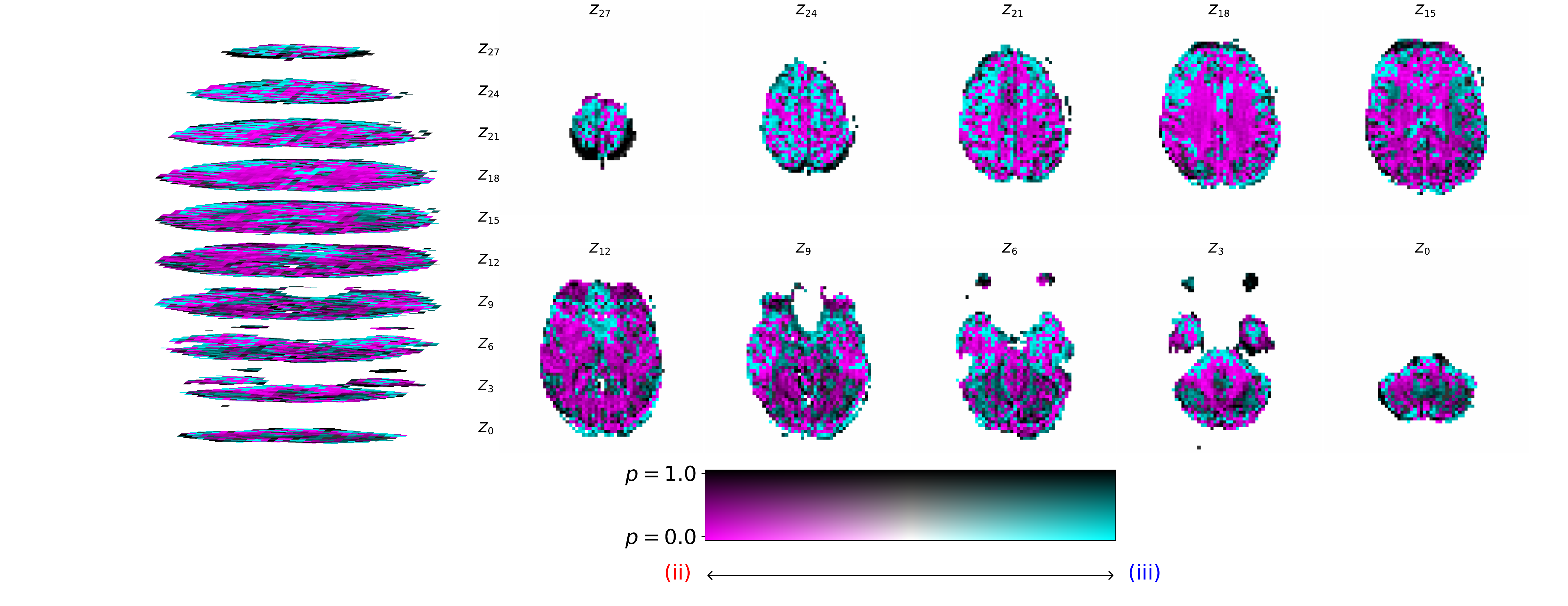}
     \end{subfigure}
     ~
     \centering
     \begin{subfigure}[t]{0.63\textwidth}
         \centering
         \includegraphics[clip, trim=5cm 9cm 5cm 13.5cm,width=\textwidth]{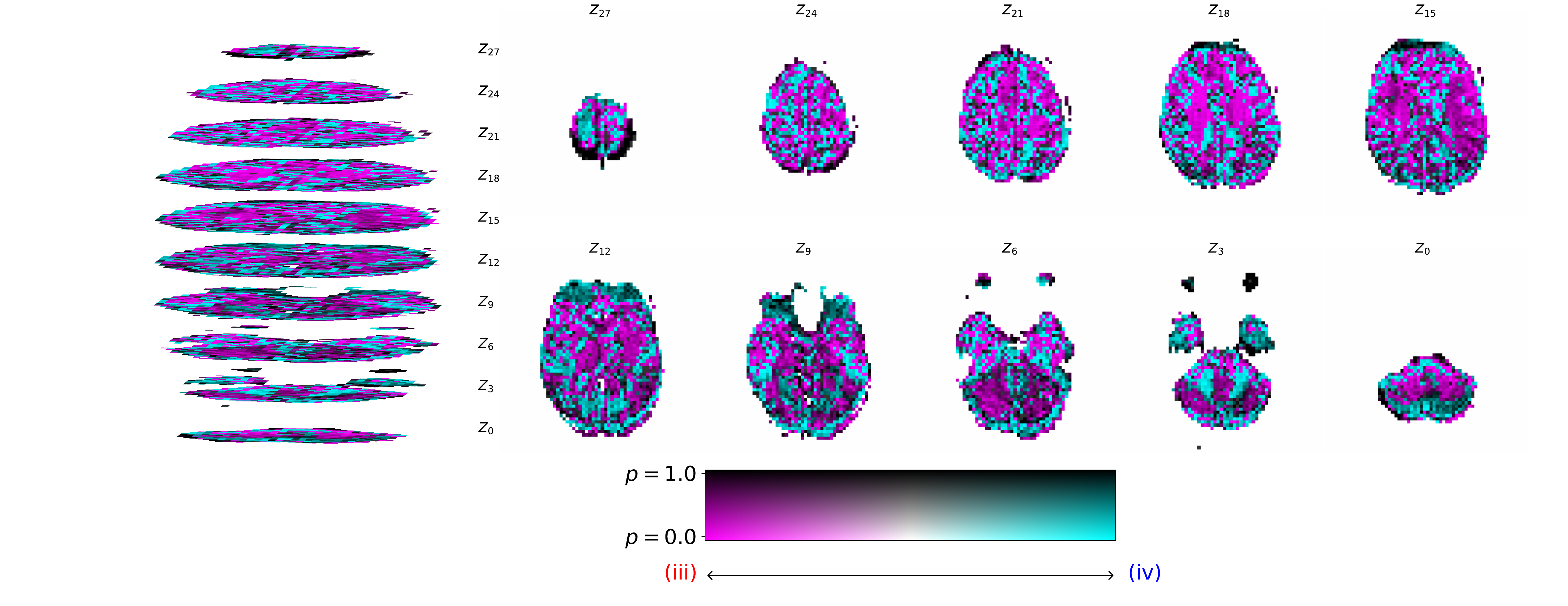}
     \end{subfigure}
    \caption{Plotting the regions of the brain synthesis ability, of each model versus the others, in NODDI dataset.}
    \label{fig:comparison_plots}
\end{figure*}

\end{document}